\documentclass[preeprint,nofootinbib,superscriptaddress,floatfix,preprintnumbers]{revtex4-1}
\pdfoutput=1
\usepackage{graphicx}
\usepackage{orcidlink}
\usepackage{hyperref}
\usepackage{amsmath}
\usepackage{amsfonts}
\usepackage{amssymb}
\usepackage{color}
\usepackage{subcaption}
\usepackage[utf8]{inputenc}
\usepackage{float}
\usepackage{microtype}
\usepackage{siunitx}
\usepackage{soul}
\usepackage[normalem]{ulem}
\usepackage[utf8]{inputenc}
\DeclareUnicodeCharacter{2212}{-}
\usepackage{comment}

\newcolumntype{L}{>{\centering\arraybackslash}m{1.99cm}}

\newcommand{\eq}{Eq.~}
\newcommand{\eqs}{Eqs.~}

\newcommand{\fig}{Fig.~}

\newcommand{\bi}{\begin{itemize}}
\newcommand{\ei}{\end{itemize}}

\begin{document}

\title{Searching for dark matter annihilating into light long-lived mediators from stars inside dwarf spheroidal galaxies}

\author{Aman Gupta \orcidlink{0000-0002-7247-2424}}\email{aman.gupta@saha.ac.in}
\affiliation{Theory Division, Saha Institute of Nuclear Physics, 1/AF, Bidhannagar, Kolkata 700064, India} 
\affiliation{Homi  Bhabha  National  Institute,  Anushakti  Nagar,  Mumbai  400094,  India}  
\author{Pooja Bhattacharjee \orcidlink{0000-0002-0258-3831}}\email{pooja.bhattacharjee@ung.si}
\affiliation{Center for Astrophysics and Cosmology, University of Nova Gorica, Vipavska 13, SI-5000 Nova Gorica, Slovenia}
\author{Pratik Majumdar \orcidlink{0000-0002-5481-5040}}\email{pratik.majumdar@saha.ac.in}
\affiliation{Theory Division, Saha Institute of Nuclear Physics, 1/AF, Bidhannagar, Kolkata 700064, India} 
\affiliation{Homi  Bhabha  National  Institute,  Anushakti  Nagar,  Mumbai  400094,  India}




\begin{abstract}
\noindent Several astrophysical and cosmological observations suggest the existence of dark matter (DM) through its gravitational effects, yet its nature remains elusive. Despite the lack of DM signals from direct detection experiments, efforts continue to focus on the indirect detection of DM from DM-rich astrophysical objects. Dwarf spheroidal galaxies (dSphs) are among the most promising targets for such searches. In this work, we aim to investigate the expected DM capture rate from the stellar component of ten nearby DM-rich dSphs, assuming that the accumulated DM eventually annihilates into light, long-lived mediators (LLLMs) which decay into gamma rays outside the dSphs. We analyze nearly 16 years of {\it Fermi}-LAT data to search for DM annihilation through LLLMs, and, from the observed stacked flux upper limits, set limits on the DM-nucleon scattering cross section for the case of a generic DM scenario. Additionally, we incorporate the Sommerfeld enhancement (SE) effect into the DM annihilation process assuming scalar DM model, and obtain bounds on the DM-nucleon scattering cross section of $\sim~10^{-36} {\rm cm}^2$ for DM masses around 100 GeV. This allows us to explore an alternative avenue for exploring DM phenomena from dSphs and compare our results with the bounds reported by direct DM detection experiments and other celestial bodies.
\vspace{0.5cm}


\end{abstract}


\maketitle

\section{Introduction}

\noindent Dwarf spheroidal galaxies (dSphs) are among the faintest, most dark matter (DM)-dominated galaxies in the Universe~\cite{Walker:2009zp, Lokas:2001mf, Battaglia:2013wqa, Walker_2013}. These small, low-luminosity systems, often found as satellites orbiting larger galaxies like the Milky Way, are characterized by an unusually high mass-to-light ratio, indicating a significant amount of DM relative to visible matter. Their lack of gas and star formation makes dSphs valuable targets for indirect DM detection, as this absence of astrophysical activity creates a relatively clean environment with minimal background noise~\cite{Battaglia:2013wqa, Strigari:2018utn}. The DM content in dSphs is inferred from the velocity dispersion of their stars~\cite{Esteban:2023xpk}, suggesting that the gravitational influence of unseen mass is substantial. Consequently, dSphs are promising candidates for searches aimed at detecting potential signals from DM annihilation or decay processes.\\
\vskip -0.4cm
\noindent In indirect detection, dSphs are particularly valuable for gamma ray searches because of their proximity and dense DM halos. Gamma rays, especially within the energy range of interest and resulting from DM annihilation, are of particular significance because they are electrically neutral, allowing them to travel directly from their source without being affected by Galactic or extragalactic magnetic fields, thus preserving information about their origin. The gamma ray space telescope like the Fermi Large Area Telescope ({\it Fermi}-LAT) has conducted extensive searches for gamma ray emissions from dSphs to constrain the DM annihilation cross section \cite{McDaniel:2023bju, Fermi-LAT:2010cni, Fermi-LAT:2011vow, Fermi-LAT:2013sme, Fermi-LAT:2015att, Fermi-LAT:2015ycq, Fermi-LAT:2016uux, Fermi-LAT:2012fij, Fermi-LAT:2015sau, Fermi-LAT:2017opo, Fermi-LAT:2016afa}. While no conclusive signals have yet been detected, the clean astrophysical environment of dSphs reduces the risk of false positives due to background contamination, making these galaxies ideal for setting stringent bounds on DM models and advancing our understanding of this elusive component of the universe.\\
\vskip -0.4cm
\noindent Most studies on DM signal from dSphs establish upper bounds on the thermally averaged DM annihilation cross section~\cite{Zhao:2017dln, Hoof:2018hyn, Lu:2017jrh, Petac:2018gue, Ando:2021jvn, Zhao:2024say}, making comparisons with direct detection challenging. This study, however, explores a distinctive mechanism: we investigate DM annihilation into light long-lived mediators (LLLMs) that decay into gamma rays outside the galaxies, allowing us to constrain the DM-nucleon scattering cross section as a function of DM mass. The minimum detectable DM mass from celestial bodies and dSphs is governed by the DM evaporation rate, setting a limit on sensitivity in both indirect and direct detection approaches. This mechanism has been developed under secluded DM models~\cite{Pospelov:2007mp, Pospelov:2008jd}, which naturally predict the existence of LLLMs~\cite{Gherghetta:2015ysa}. Theoretically, such particles are well motivated~\cite{Holdom:1986eq,Martin:1997ns, Batell:2009zp,Bell:2021pyy} and are actively being searched for in collider and fixed-target experiments~\cite{Reece:2009un,Morrissey:2014yma,ATLAS:2015oan,ATLAS:2023cjw}. Numerous works have considered LLLM scenarios in indirect detection (e.g., \cite{Batell:2009zp, Schuster:2009au, Leane:2017vag, Cermeno:2018qgu, Bell:2021pyy, Andrade:2024ekx, HAWC:2018szf, Arina:2017sng, Niblaeus:2019gjk, Bell:2019pyc, Dasgupta:2020dik}). In particular, LLLMs have been invoked to explain the observed excesses in cosmic-ray electrons and positrons, as discussed in Refs.~\cite{Rothstein:2009pm,Arkani-Hamed:2008hhe,Kim:2017qaw}. Ref.~\cite{Pospelov:2008jd} further shows that LLLM scenarios can enhance the DM annihilation rate, leading to stronger indirect detection signals. Various astrophysical objects have been considered as potential probes of secluded DM and LLLMs, including the Sun~\cite{Leane:2017vag,HAWC:2018szf,Andrade:2024ekx}, Jupiter~\cite{Leane:2021tjj,Li:2022wix}, the Earth~\cite{Feng:2015hja}, as well as brown dwarfs (BDs) and neutron stars (NSs)~\cite{Leane:2021ihh, Bhattacharjee:2022lts}. Yet our approach combines this with the inclusion of the Sommerfeld effect, which enhances annihilation rates at low velocities, providing a boost to the DM signal. Sommerfeld enhancement (SE) is the widely discussed phenomenon in the context of DM annihilation~\cite{Arkani-Hamed:2008hhe, Petac:2018gue, Beneke:2022rjv, Wang:2023wbw, Phoroutan-Mehr:2024cwd}. The authors in~\cite{Lu:2017jrh} explored the effects of SE on the DM annihilation cross section in dSphs using {\it Fermi}-LAT data. Ref.~\cite{Beneke:2022rjv}, for instance, considers a combined effect of resonant annihilation and SE in the Standard Model (SM) Higgs portal and MSSM-inspired DM scenarios.\\
\noindent In this work, to the best of our knowledge, for the first time, we investigate the expected DM capture rate within the stellar component of ten nearby dSphs, hypothesizing that captured DM annihilates into  LLLMs. Later we also incorporate the Sommerfeld effect in this framework considering a particular scalar DM model interacting with the scalar mediator \footnote{ In the first part of this work, we derive bounds on DM-nucleon scattering cross section without assuming any specific particle nature of DM and thus refer to these bounds as ``model independent". Whereas, in the second part, we incorporate the Sommerfeld Enhancement (SE), which depends on the details of the DM model. For this (SE) case, we consider a scalar DM model, making our results ``model dependent". Please note that in both cases (with and without SE) the mediator is taken to be a scalar particle, which subsequently decays into gamma photons to produce observable signals.}. Using nearly 16 years of {\it Fermi}-LAT gamma ray data, we set upper limits on the gamma ray flux and refine constraints on the DM-nucleon scattering cross section. This complementary method broadens traditional indirect detection approaches by focusing on stellar capture and mediator decay scenarios, offering a new avenue for probing DM interactions in dSphs.\\
\noindent We organize the paper in the following manner. Section \ref{sec:dsph_target} begins with a brief discussion about the choice of our selected dSphs, listing their important properties used in this work. Section \ref{sec:fermi_data} deals with the {\it Fermi}-LAT data analysis of the selected dSphs and calculation of gamma ray flux upper limits. The formalism for DM capture and annihilation via LLLM from dSphs and the estimation of the gamma ray spectrum from such processes have been reviewed in section \ref{sec:dm_capture} and section \ref{sec:llm_flux}. In section \ref{sec:fermi_bound}, we derive the constraints on DM-nucleon scattering cross section using {\it Fermi}-LAT observational data.  Section \ref{sec:sommerfeld} explores the simple model of the SE along with the decay of the LLLM, $\phi$, into gamma rays. Finally, in section \ref{sec:conclusion}, we conclude with a summary and prospect for future studies. 
\section{Our Targets: Nearby dSphs}
\label{sec:dsph_target}
\begin{table}[hbt]
    \centering 
    \caption{Properties of our selected dSphs. Please see the texts for more details.}

    \begin{tabular}{|c|c|c|c|c|c|c|}
    \hline
    Source & RA [deg] & DEC [deg] & d (kpc) &  $R_{\ast}$ (pc)  &$\sigma_{\rm l.o.s}$ (km/s) & $M_{\ast, \rm tot}$ ($M_\odot$)\\
        \hline
         Draco II & 238.17 & 64.58 & $21.57^{+50}_{-0.49}$ & 13.03 & $<5.9$ & 30 \cite{Moskowitz:2019imu}\\
         \hline
         Segue I & 151.75 & 16.08& $22.90^{+2.21}_{-2.01}$  & 15.33 & $3.7^{+1.4}_{-1.1}$ & 220 \cite{Moskowitz:2019imu}\\
         \hline
         Sagittarius & 283.83 &-30.55 & $26.30^{+1.88}_{-1.75}$ & 1199.45 & $11.4^{+0.7}_{-0.7}$ & 21$\times 10^6$ \cite{McConnachie_2012}\\
         \hline
         Hydrus I& 37.39 &-79.31 & $27.54^{+0.51}_{-0.50}$  & 40.62 & $2.7^{+0.51}_{-0.43}$ &3.0 \cite{Moskowitz:2019imu}\\
         \hline
         Reticulum II & 53.92 &-54.05 & $31.62^{+1.49}_{-1.42}$ & 23.76 & $3.6^{+1}_{-0.7}$ &764.0 \cite{Moskowitz:2019imu}\\
         \hline
         Ursa Major II & 132.87 & 63.13 & $34.67^{+2.13}_{-2.01}$ & 65.15 & $6.7^{+1.4}_{-1.4}$ & 296.0 \cite{Moskowitz:2019imu}\\
         \hline
         Carina II & 114.11 & -58.0 & $37.39^{+0.39}_{-0.39}$ & 59.01 & $3.4^{+1.2}_{-0.8}$ &0.38 $\times 10^6$ \cite{McConnachie_2012}\\
         \hline
         Bootes II& 209.51 & 12.86 & $41.68^{+1.16}_{-1.13}$ & 29.89 & $2.9^{+1.6}_{-1.2}$ &298.0 \cite{Moskowitz:2019imu}\\
         \hline
        Willman I& 162.34 & 51.05 &$38.01^{+7.68}_{-6.39}$ & 15.33 & $4^{+0.8}_{-0.8}$ &455.0 \cite{Moskowitz:2019imu}\\
         \hline
        Coma Berenices& 186.75 & 23.91 & $42.26^{+1.58}_{-1.52}$ & 43.68 & $4.6^{+0.8}_{-0.8}$ & 1307.0 \cite{Moskowitz:2019imu}\\
         \hline
    \end{tabular}
\label{tab:dsphs_list}
\end{table}

\noindent The Milky Way hosts a large population of dSphs, which are small, faint, and have minimal star formation activity compared to larger galaxies. These dSphs are highly DM-dominated, evidenced by their high mass-to-light ratios, and often embedded within extended DM halos. Their simple structure and relatively low baryonic content make dSphs ideal laboratories for studying DM properties because the effects of DM are less contaminated by stellar and gas dynamics. Given their proximity and abundance in the Milky Way, these halos provide an accessible means to examine DM interactions, including capture rates by stars within dSphs. By analyzing how DM might accumulate in stars, we can infer properties about DM particle interactions, helping to constrain models of DM capture and potentially shine a light on the nature of DM particles.

\noindent In this study, we are interested in assessing the potentiality of dSphs in constraining the DM parameter space under the mechanism where DM annihilates through the decay of intermediate particles, as mentioned above. In this regard, we restrict ourselves to only those dSphs whose distance from the Milky Way is less than 50 kpc which corresponds to only 10 nearby dSphs~(Table~\ref{tab:dsphs_list}), as the flux falls with distance ($\propto 1/d^2$).

\noindent In Table~\ref{tab:dsphs_list}, we describe the characteristic of our selected dSphs as follows: columns I \& II: right ascension (RA) and declination (DEC) of our targets in degree; column III: heliocentric distance (d) in kpc;  column IV: stellar radius ($R_\star$) in pc; column V: velocity dispersion ($\sigma_{\rm l.o.s}$) in km/s and  column VI: total stellar mass ($M_{\star,{\rm tot}}$) in unit $M_\odot$. We calculate the stellar radius $R_\star$ and $M_{\star,{\rm tot}}$ in Sec. \ref{sec:dm_capture} following Ref.~\cite{Moskowitz:2019imu}. To estimate $M_{\star,{\rm tot}}$ and $R_\star$, we take the values of $N_{\rm tot}$ from Table 3 of~\cite{Moskowitz:2019imu} except for Sagittarius (Sgr) and Carina II cases for which we use the approximate values of total stellar mass reported in~\cite{McConnachie_2012}. The values of other parameters such as RA, DEC, d, and $\sigma_{\rm l.o.s}$ are taken from~\cite{pace2024, pace2025dSphs} unless indicated in Table~\ref{tab:dsphs_list}. This Table also mentions the astrophysical uncertainties corresponding to the relevant dSph parameters $d$ and $\sigma_{\rm l.o.s}$.

\section{{\it Fermi}-LAT data analysis for dSphs}
\label{sec:fermi_data}
\subsection{Data selection}

\noindent We analyze nearly 16 years of {\it Fermi}-LAT data, spanning from August 4, 2008, to June 2, 2024. For our analysis, we utilize \texttt{Fermipy} version 1.1.0 and \texttt{Fermi ScienceTools} version 2.2.0\footnote{\url{https://fermi.gsfc.nasa.gov/ssc/data/analysis/software/}}. The data is processed with the source class instrument response function $\rm{P8R3\_SOURCE\_V3}$\footnote{\url{https://fermi.gsfc.nasa.gov/ssc/data/analysis/documentation/Pass8_usage.html}}.

\subsection{Analysis technique and gamma ray flux upper limits}

\noindent We focus on the energy range $E \in [0.5, 500]$~GeV and extract data within a $15^{\circ}$ region of interest (ROI) centered on each dSph location. Our ``source model'' includes the ``source of interest'' along with all sources within the $15^{\circ}$ ROI from the 4FGL-DR4 catalog\cite{Fermi-LAT:2022byn}\footnote{\url{https://fermi.gsfc.nasa.gov/ssc/data/access/lat/14yr_catalog/}}. Additionally, we incorporate the galactic diffuse model ($\rm{gll\_iem\_v07.fits}$) and the isotropic diffuse model ($\rm{iso\_P8R3\_SOURCE\_V3\_v1.txt}$)\footnote{\url{https://fermi.gsfc.nasa.gov/ssc/data/access/lat/BackgroundModels.html}}. After constructing the source model and generating all necessary input files, we perform a bin-by-bin binned likelihood analysis\footnote{\url{https://fermi.gsfc.nasa.gov/ssc/data/analysis/scitools/binned_likelihood_tutorial.html}} on the extracted data. During this analysis, the spectral parameters of all sources within a $10^{\circ} \times 10^{\circ}$ ROI and the normalization parameters for the two diffuse background models are allowed to vary freely.

\noindent We model the gamma ray emission from each dSph as a new pointlike source, assuming a power-law spectrum, ${\rm d}N/{\rm d}E \propto E^{-\Gamma}$, with a spectral index $\Gamma = 2$. To search for evidence of excess emission from the dSph location, we calculate the test statistic (TS), defined as TS$= -2\ln(L_{\rm max,0}/L_{\rm max,1})$, where $L_{\rm max,0}$ and $L_{\rm max,1}$ are the maximum likelihood values for the null hypothesis (background only) and the alternative hypothesis (including the additional source), respectively. No significant excess emission is detected at any dSph location, with TS values falling well below the point source detection threshold of TS$=25$.

\noindent In the absence of excess emission, we calculate the 95\% confidence level (CL) upper limits on the gamma ray flux for each dSph using the profile likelihood method~\cite{Rolke:2004mj}. In this approach, the fit is performed until the difference in the log-likelihood function, $-2\Delta\ln(\mathcal{L})$, reaches 2.71, corresponding to the one-sided 95\% CL.

\begin{figure} [hbt]  
\hspace{-0.5cm}
\includegraphics 
[width=18.4cm,height=25.5cm]{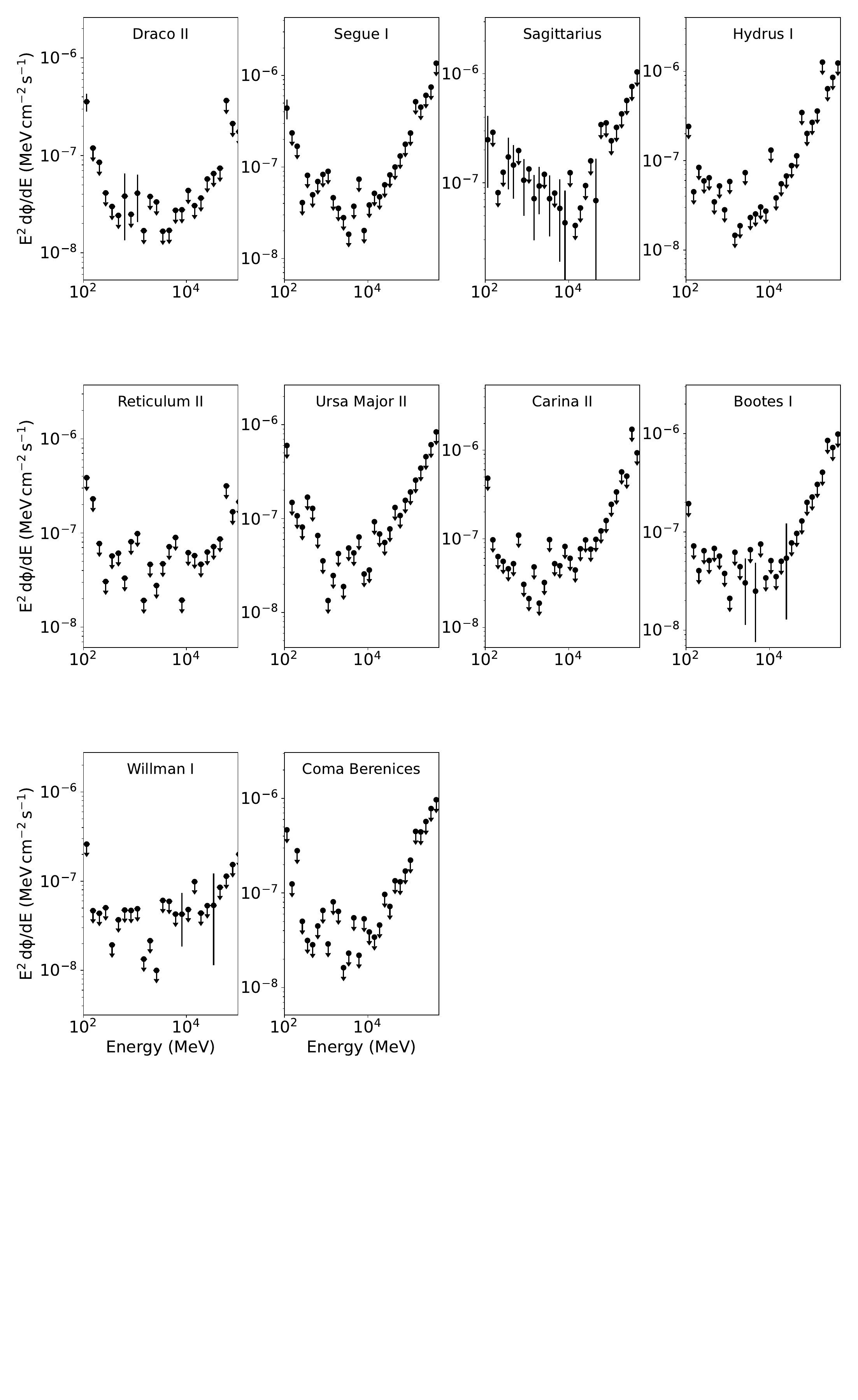}
\vspace{-5.5cm}
\caption{Bin-by-bin flux upper limits at $95\%$ CL observed by {\it Fermi}-LAT for our selected dSphs.} 
\label{fig:diff_flux_fermi}
\end{figure}

\noindent In Fig.~\ref{fig:diff_flux_fermi}, we show the observed bin-by-bin $E^2~{\rm d}\Phi/{\rm d}E$ flux upper limits for all ten dSphs considered in this work. 

\section{Formalism for gamma ray flux calculation via LLLMs from stars inside dSphs}
\label{sec:dm_capture}
In this section, we present the key formulas and assumptions necessary to calculate the gamma ray flux resulting from the two-step cascade annihilation of DM particles. We consider a scenario in which DM is captured by stellar components of dSphs in sufficient quantities, primarily through the scattering of nucleons. This interaction causes DM to lose kinetic energy and become gravitationally bound to the dSph. If accumulated in considerable amounts, such DM particles undergo self-annihilation producing gamma rays. It should be noted that in the two-step annihilation process, the DM particles first annihilate into LLLMs which later decay into gamma rays.  

\subsection{DM capture inside the stars of dSphs}

\noindent DM, when traversing through celestial objects, undergoes single or multiple scattering with nucleons depending on the kinetic energy of DM. In due time, this process leads to their thermalization and eventual capture when their velocity falls below the escape velocity of dSph. 

\noindent It should be noted that, in dense stellar environments stars such as neutron stars or white dwarfs, the stellar density, \( \rho_{\ast}(r) \), is extremely high.  In contrast, dSphs present a different scenario, as the majority of their mass resides in DM rather than baryons. Thus, to treat dSph, we need to account for the fact that: (i) the dSph galaxies have a very low baryon-to-DM ratio, and (ii) the nucleon density in dSphs is primarily from old stars with minimal contributions from gas or dust.  These factors may produce enough DM interactions in the dSphs population, despite significantly smaller star counts. The stellar density can be modeled using a Plummer profile:
\begin{align}
   \rho_{\ast}(r) = \frac{3 M_{\ast,\rm tot}}{4 \pi R_\ast^3} \left( 1 + \frac{r^2}{R_\ast^2} \right)^{-5/2}
   \label{eq:plummer}
\end{align}

\noindent where \( M_{\ast,\rm tot} \)~\cite{Moskowitz:2019imu} is the stellar mass, expected to be \(\sim\) \( N_{\rm tot}~\times~M_{\odot} \)  and \( R_\ast \) is the stellar radius of each star related to the half-light radius (\( R_h \)) as \( R_\ast = \sqrt{2^{2/3}-1} R_h \). \(N_{\rm tot}\) represents the total expected number of stars in each dSph, determined by the stellar density profile of dwarf spheroidal galaxies. For the Plummer profile mentioned in \eq \ref{eq:plummer} we adopt values from recent studies~\cite{Moskowitz:2019imu, Mu_oz_2018}, where the number of member stars was derived by fitting the Plummer density profile to observational data. Alternatively, the star counts can be approximated using the luminosities of each dSph, as utilized in~\cite{Bogorad:2024hfj}. We verify that both methods yield consistent results, with no significant impact on our conclusions. The maximum capture rate of DM particles by the stars of the dSphs is
given by \cite{Bernal:2012qh, Bottino:2002pd,Garani:2017jcj, Leane:2020wob}

\begin{align}
\label{eq:cmax}
    C_{\rm max} = \pi R_{\ast}^2 n_\chi(r) v_0\left(1 + \frac{3}{2}\frac{v_{\rm esc}^2}{v_d(r)^2}\right) \,
\end{align}
\noindent The number density of DM particles, $n_\chi(r)$, at a distance r from the Galactic Center is $n_\chi(r) =  \rho_\chi(r)/m_\chi$, where $m_\chi$ is DM mass and $\rho_\chi(r)$ is DM density. The velocity dispersion of the DM halo $v_d(r)$ is related to the orbital velocity $v_c(r)$ at distance $r$ by $v_d^2(r) = \frac{3}{2}v_c^2(r)$, with $v_c^2(r) = \frac{G M(r)}{r}$, where $G$ denotes the universal gravitational constant and $M(r)$ the mass of dSph within a radius of $r$. The average speed in the DM rest frame $v_0$ is computed using the relation $v_0^2 = \frac{8}{3\pi} v_d^2(r)$ and for escape velocity we use $v^2_{\rm esc} = \frac{2 G M_{1/2}}{R_{1/2}}$, where $M_{1/2}$ is the expected mass of dSphs contained within half-light radius, $R_{1/2}$. For the case of dSphs we fix $r= R_{1/2}, M(r) = M_{1/2}$. It should be noted that $M_{1/2}$ approximately can be expressed in terms of velocity dispersion of dSphs, $\sigma_{\rm l.o.s}$, as  $M_{1/2}~=~\frac{2.5}{G}\sigma^2_{\rm l.o.s} R_{1/2}$ \cite{Bhattacharjee:2020phk}.\\

\noindent To model the distribution of DM, we select the widely recognized Navarro-Frenk-White (NFW) density profile \cite{Navarro:1995iw, Navarro:1996gj}. The NFW profile is particularly suited for describing DM halos in galaxies and clusters, capturing their density structure on a large scale. The general form of the NFW profile is given by: 
\begin{align}
    \label{eq:nfw}
    \rho^{\rm NFW}_\chi(r) = \rho_s\frac{r_s}{r}\left( 1 + \frac{r}{r_s}\right)^{-2}\,
\end{align}

\noindent where $\rho_s$ and $r_s$ represent the characteristic density and scale radius, respectively. In the present analysis, we use the analytical formulas \footnote{It is worthwhile to mention here that though for the NFW parameters, namely, $\rho_s$ and $r_s$, we use the above-mentioned expressions, we have also calculated the gamma ray flux adopting the $\rho_s$ and $r_s$ values directly from the various Refs. \cite{Geringer-Sameth:2014yza, Hu:2023iex, Calore:2022stf, LHAASO:2024upb, Acharyya:2024tvg} and compared with those obtained from the analytical expressions and we find no significant changes in the main results.} for $r_s$ and $\rho_s$ taken from Refs.~\cite{Bhattacharjee:2020phk, Evans:2016xwx}.

\noindent In our investigation, in addition to single scattering, we also incorporate multiple scattering of DM particles as for massive DM particles, the energy loss in one collision may not be enough for gravitational capture. The inclusion of multiple collisions modifies the total capture rate of \eq \eqref{eq:cmax} as a series given by
\begin{align}
    \label{eq: ctot}
    C_{\rm tot}(r) =  \sum_{n=1}^{\infty} C_{\rm n}(r)\,, 
\end{align}

\noindent where $C_{\rm n}$ denotes the capture rate corresponding to the ``n'' number of collisions after which the DM velocity falls below the escape velocity and eventually gets captured by dSph. The approximate formula for $C_{\rm n}$ is adopted from Refs.~\cite{Bramante:2017xlb,Ilie:2020vec} and is written as

\begin{align}
    \label{eq:Cn}
    C_{\rm n}(r) = \pi R_{1/2}^2\mathcal{P}_n(\tau)\frac{\sqrt{6}n_\chi(r)}{3\sqrt{\pi}v_d(r)}\left[(2v_d(r)^2 + 3v_{\rm esc}^2) - (2v_d(r)^2 + 3v_{\rm n}^2)\exp\left(-\frac{3(v_{\rm n}^2 - v_{\rm esc}^2)}{2v_d(r)^2} \right)\right]\,,
\end{align}
where $v_{\rm n} = v_{\rm esc}\left(1 - \frac{2m_\chi m_{\rm n}}{(m_\chi +m_{\rm n})^2} \right)^{-{\rm n}/2}$, with $m_\chi$ and $m_{\rm n}$ denoting the DM mass and mass of the nucleon, respectively. The probability ($\mathcal{P}_{\rm n}(\tau)$) of the DM particles with optical depth $\tau$ to collide exactly ${\rm n}$ times before they get captured by the dSph can be written as 
\begin{align}
    \label{eq:prob}
\mathcal{P}_{\rm n}(\tau) = 2\int_{0}^{1} \frac{z e^{-z\tau}(z\tau)^{\rm n}}{{\rm n}!} \,dz . 
\end{align}

\noindent The optical depth $\tau$ is defined in terms of DM-nucleon scattering cross  section ($\sigma_{\chi n}$) as $\tau = \frac{3\sigma_{\chi n} N_T}{2\pi R_{\ast}^2}$, where $N_T = \frac{M_{\ast,\rm tot}}{m_n}$ is the total number of nucleons in the target.

\subsection{Gamma ray spectrum from DM annihilation via LLLMs}

\noindent The captured DM particles may undergo self-annihilation if their accumulation inside dSph is sufficient. At time $t$, the evolution of the total number of captured DM particles $N(t)$ can be expressed as

\begin{align}
    \label{eq:dndt}
    \frac{dN(t)}{dt} = C_{\rm C} - C_{\rm E}N(t) - C_{\rm{ann}}N^2(t)\,
\end{align}

\noindent where $C_{\rm C}$ is the total capture rate, which is redefined as $C_{\rm C} = {\rm min}[C_{\rm tot}, C_{\rm max}]$ to properly incorporate the perturbative estimation, especially relevant for multiple scattering, while $C_{\rm E}$ denotes the rate at which the captured DM evaporates by scattering. In \eq \eqref{eq:dndt}, $C_{\rm ann} = \langle\sigma_{\rm ann} v\rangle/V_0$ represents the annihilation rate with $\langle\sigma_{\rm ann} v\rangle$ and $V_0$ being the velocity averaged annihilation cross section and volume over which annihilation occurs. We find that the evaporation mass ($m_{\rm evp}$) of a typical dSph is around $\sim 8~{\rm GeV}$ \cite{Garani:2021feo} which is higher than the $m_{\rm evp}$ of Sun, i.e.,  $m_{\chi}\geq4$ GeV \cite{1987ApJ...321..560G, Busoni:2013kaa}. It is important to note that in the present analysis, we only consider the DM mass $m_\chi \gtrapprox 10~{\rm GeV}$ and for such higher DM masses, we can safely disregard the contribution of DM evaporation effect due to scattering ($C_{\rm E}$) in our calculation. Assuming the initial condition as $N(0) = 0$ at $t = 0$, the general solution of \eq \eqref{eq:dndt} is given by

\begin{align}
    \label{eq:N_total}
    N(t) = C_{\rm C}\, t_{\rm eq}\,\tanh (t/t_{\rm eq})\, ,
\end{align}
\noindent where $t_{\rm eq}$ is the time required to reach equilibrium between DM capture and annihilation and is defined as 
\begin{align}
    \label{eq:teq}
    t_{\rm eq} = \frac{1}{\sqrt{C_{\rm C}\,C_{\rm ann}}} .
\end{align}
\noindent The dSphs are primarily composed of old, low-mass stellar populations. The majority of stars in dSphs are ancient, typically over $10^{9}$ years old, having formed early in the universe's history. These stars are metal poor because star formation in dSphs ceased relatively early, limiting the enrichment of heavy elements. 
The equilibrium timescale, $t_{\rm eq}$, between DM annihilation and capture rate depends on the core density ($\rho_{\star,c}$) and core temperature ($T_{\star,c}$) of stars inside dSphs where annihilation takes place. In the case of dSph stars, accurately estimating these stellar properties is particularly challenging due to the limited availability of high-resolution photometric and spectroscopic studies. Consequently, it remains uncertain whether these systems have reached equilibrium. However, adopting a conservative approach, we assume equilibrium is achieved. Under this assumption, the total annihilation rate ($\Gamma_{\rm ann}$) of DM particles can be expressed as

\begin{align}
    \label{eq:Ann_rate}
    \Gamma_{\rm ann} = \frac{C_{\rm ann}\,N^2}{2}\rightarrow \frac{C_{\rm C}}{2}\, ,
\end{align}
\noindent where a factor of $2$ indicates that in each self-annihilation process, two DM particles participate. After computing the total annihilation rate, we can express the expected flux ($E^2\,\frac{d\Phi}{dE}_{\rm exp}$) from DM annihilation for LLL mediators and later compare that with the observed differential flux of gamma rays at the {\it Fermi}-LAT detector by the following equations,
\begin{align}
    \label{eq:dphidE_exp}
    E^2\,\frac{d\Phi}{dE}_{\rm exp}  = \frac{\Gamma_{\rm ann}}{4\pi\, d^2}\times E^2\,\frac{dN_\gamma}{dE}\, ,
\end{align}   

\begin{align}
    \label{eq:dphidE_fermi}
    E^2\,\frac{d\Phi}{dE}_{\rm exp} = E^2\,\frac{d\Phi_\gamma}{dE}_{\rm {\it Fermi}-LAT}\, ,
\end{align}

\noindent where $\frac{dN_\gamma}{dE}$ is the gamma ray spectrum from dSph and $d$ denotes the heliocentric distance of dSph. In this work, we consider a model in which the spectrum $dN_\gamma/dE$ originates from the decay of the mediators ($\chi\chi\rightarrow\phi\phi; \phi\rightarrow \gamma\gamma$) unlike the case where DM directly annihilates to gamma rays ($\chi\chi\rightarrow\gamma\gamma$). The two-step process has an advantage over the annihilation of DM particles directly to SM states as shown in Ref. \cite{Pospelov:2008jd}. In scenarios where DM directly annihilates into SM particles within astrophysical objects, the resulting flux of SM particles may be reduced due to further trapping within the object. However, if DM instead annihilates into SM particles through LLLMs ($\phi$) that can escape the dSphs, the observable flux could be enhanced, increasing the potential for detection. Additionally, in this latter scenario, four photons are produced per annihilation, further boosting the gamma ray flux compared to direct annihilation. Note that following Refs. \cite{Leane:2021ihh, Bhattacharjee:2022lts}, we also here ignore any possible interactions of the mediator and SM particles inside the dSphs and assume all the mediators decay outside the dSph yielding detectable gamma ray flux. This two-step mechanism also assumes that the decay length of the mediator $L_\phi = \frac{m_\chi}{m_\phi\Gamma}$, with $\Gamma$ being the decay width, is much larger than the half-life radius of dSph $R_{1/2}$. This automatically implies that the mediator should be ``light" i.e. $m_\chi\gg m_\phi$ and ``long-lived" i.e., the decay width should be very small or the mediator's lifetime ($\tau_\phi$) should be considerably long. This is indeed the case for the secluded DM models \cite{Pospelov:2007mp,Arkani-Hamed:2008hhe,Pospelov:2008jd}. One can get an estimate of $m_\phi$ and $\tau_\phi$ by using the fact that $\phi$ should decay outside the dSphs, i.e.,
\begin{align}
    L_\phi = \gamma_\phi c\tau_\phi \gg R_{1/2}\,,
    \label{eq:Lphi}
\end{align}
where $\gamma_\phi = m_\chi/m_\phi $ is the boost factor and $c$ denotes the speed of light in vacuum. Considering a half-light radius $R_{1/2} = 100$ pc and for DM mass in GeV-TeV scale, the typical values of the mediator lifetime and mass are $\tau_\phi \approx 10^{10}$ s and $m_\phi\ll\mathcal{O}(1)$ MeV (using $m_\chi/m_\phi\gg 10^{10}/\tau_\phi)$. Mediators with such long lifetimes, particularly for indirect detection of DM, have been discussed in previous works; for instance, the positron excess has been explained using LLLMs in Refs.~\cite{Rothstein:2009pm,Kim:2017qaw}. The authors in Ref.~\cite{Gori:2018lem} have studied DM searches from dSphs from the decay of such LLLMs. From the cosmological point of view, such long-lived mediators are possible if their abundance is small enough during the early Universe. Several mechanisms to accommodate such scenarios are explored in Refs.~\cite{Rothstein:2009pm,Kim:2017qaw,DEramo:2018khz}. Here, we mainly focus on the observations of DM annihilation via such LLLMs. Since, in the present work, we are considering DM mass $m_\chi \ge 10$ GeV, so for $m_\phi < \mathcal{O}(1)$ MeV the condition $m_\chi\gg m_\phi$ is satisfied all of the time.  \\  

\noindent With the assumptions mentioned above, the gamma ray spectrum originating from the DM annihilation from dSphs through an LLLM is then given by a box-shaped spectrum described in \cite{Ibarra:2012dw} and can be written as
\begin{align}
  \frac{dN_\gamma}{dE} =  \frac{4}{\Delta E}\Theta\left(E - E_{-}\right)\Theta\left(E_{+} - E\right)\, 
  \label{eq:diff_flux}
\end{align}
\noindent where $\Theta$ denotes the usual Heaviside function while the upper and lower limits of gamma ray energy are written as $E_{\pm} = \frac{1}{2}\left(m_\chi \pm \sqrt{m_\chi^2-m_\phi^2} \right)$ which are also referred to the right (+ sign) and left ($-$ sign) edges of the box. The width of the box is defined as $\Delta E = E_{+} - E_{-} = \sqrt{m_\chi^2 -m_\phi^2}$, while the center is given by $E_0 = (E_{+} + E_{-})/2 = m_\chi/2$. \\


\section{Expected flux from DM annihilating to LLLMs from dSphs}
\label{sec:llm_flux}
\begin{figure}[hbt]
\hspace*{-1.5cm}
\includegraphics[width=8.8cm,height=7.5cm]{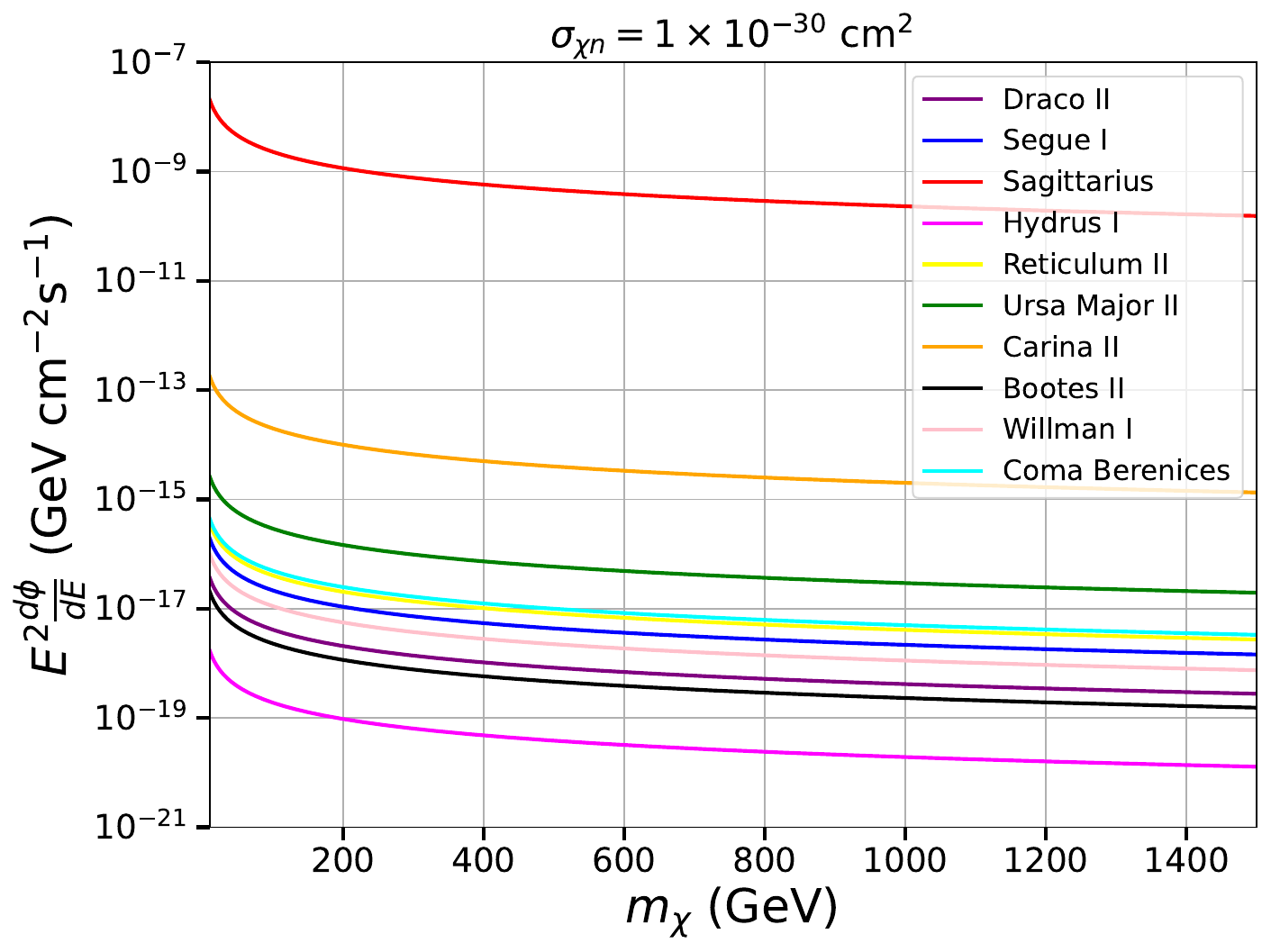}
\hspace*{-0.01cm}
\includegraphics[width=8.8cm,height=7.5cm]
{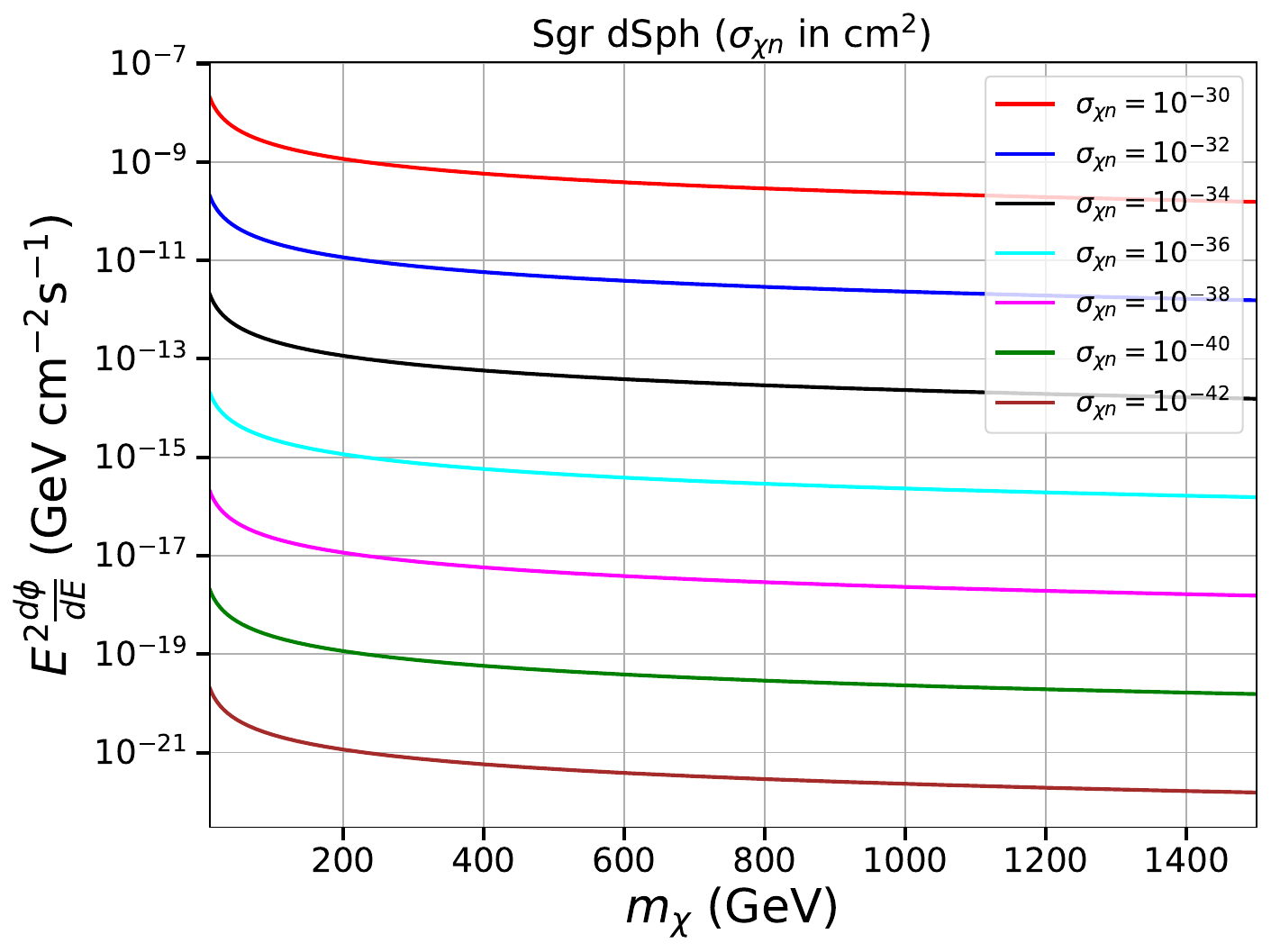}
    \caption{Differential gamma ray flux as a function of DM mass ($m_\chi$) for ten different dSphs. The left (right) plot corresponds to a fixed (different) value of DM-nucleon scattering cross section($\sigma_{\chi n}$).}
    \label{fig:flux_plot}
\end{figure}

\noindent In this section, we study the expected flux from our targets following Eq.~\ref{eq:dphidE_exp} for $m_\chi\gg m_\phi$. Fig. \ref{fig:flux_plot} illustrates the variation of gamma ray differential flux from DM annihilation as a function of DM mass for our selected dSphs. In Table~\ref{tab:list_nfw}, we mention the values (and their associated uncertainties) of $R_{1/2}, M_{1/2}$, $r_s$  and $\rho_s$ for all of our targets, which are crucial to derive the flux limits presented in Fig.~\ref{fig:flux_plot}. Here we would like to note that, throughout this paper, we adopt the central values of the parameter sets (see Tables \ref{tab:dsphs_list} and \ref{tab:list_nfw}) to derive limits from the selected dSphs and in Sec.~\ref{sec:uncertainty}, we discuss how the uncertainty on parameter space can propagate through our obtained limits. The range of DM mass is taken from $10$ GeV to $1500$ GeV, and a typical value of the DM nucleon scattering cross section, $\sigma_{\chi n} = 1\times 10^{-30}~\rm cm^{2}$, is adopted. It should be mentioned that, as long as the condition $m_\chi\gg m_\phi$ holds, the DM-induced gamma ray flux limits are more or less independent of the mediator mass $m_\phi$.  With increasing mediator mass, the width of the spectrum changes, and the flux is shifted towards higher DM masses. \\ 

\noindent From the left panel of Fig.~\ref{fig:flux_plot}, we observe that for the Sagittarius (Sgr) dSph, the expected gamma ray flux is stronger compared to other chosen dSphs. This is because Sgr has significantly higher $M_{\ast, \rm tot}$ and $R_{\ast}$, leading to increased DM accumulation and, consequently, a higher DM annihilation rate compared to other dSphs such as Draco II and Segue I. Therefore, we expect the Sgr bounds on DM parameter space to be stronger than other dSphs. The right plot of Fig.~\ref{fig:flux_plot} displays how the gamma ray flux from Sgr dSph varies with different values of $\sigma_{\chi n}$. We show the expected behavior of flux, i.e., the flux decreases with the DM-nucleon scattering cross section. This occurs because higher scattering cross sections result in more efficient capture of DM by the stellar component, reducing the available annihilation rate and, hence, the resulting gamma-ray flux.
\begin{table}[hbt!]
\centering
\caption{Sample of dSphs used in this study with the median value of their associated NFW density profile parameters. column I: $M_{1/2}$ in unit $M_\odot$ \cite{pace2024, pace2025dSphs}; column II: $R_{1/2}$ in (pc) \cite{pace2024, pace2025dSphs}; column III: $\rho_s$ in unit $\rm GeV/cm^{3}$ and column IV: $r_s$ in unit kpc. The values of $\rho_s$ and $r_s$ are computed following the expressions given in \cite{Bhattacharjee:2020phk, Evans:2016xwx}.}
\label{tab:list_nfw}
\begin{tabular}{|c|c|c|c|c|}
\hline
Source & $M_{1/2}$($M_\odot$) &$R_{1/2}$ (pc)  &$\rho_s$ ($\rm GeV/cm^{3}$) & $r_s$ (kpc) \\
\hline
Draco II & $<5.34\times 10^{5}$ & $16.01^{+3.60}_{-3.51}$ & $<~44.13^{+30.19}_{-13.57}$  & $0.08^{+0.02}_{-0.02}$\\
\hline
Segue I & $2.45^{+2.06}_{-1.39}\times 10^{5}$ & $19.51^{+3.38}_{-3.13}$ & $11.69^{+20.45}_{-7.41}$  & $0.09^{+0.01}_{-0.01}$\\
\hline
Sagittarius & $1.88^{+0.30}_{-0.26}\times 10^{8}$ &  $1565.80^{+133.09}_{-126.16}$ & $0.02^{+0.01}_{-0.01}$  & $7.83^{+0.66}_{-0.63}$\\
\hline
Hydrus I  & $3.50^{+1.43}_{-1.13}\times 10^{5}$ & $52.42^{+5.33}_{-5.55}$  & $0.86^{+0.67}_{-0.36}$  & $0.26^{+0.03}_{-0.03}$\\
\hline
Reticulum II  & $4.30^{+2.46}_{-1.85}\times 10^{5}$ & $36.46^{+5.25}_{-5.61}$ & $3.17^{+4.18}_{-1.57}$  & $0.18^{+0.03}_{-0.03}$ \\
\hline
Ursa Major II  & $3.83^{+1.82}_{-1.44}\times 10^{6}$ & $92.06^{+7.40}_{-7.03}$ & $1.72^{+1.24}_{-0.79}$  & $0.46^{+0.03}_{-0.03}$\\
\hline
Carina II  & $8.12^{+5.79}_{-4.06}\times 10^{5}$ & $76.51^{+7.69}_{-7.72}$ & $0.64^{+0.82}_{-0.33}$  & $0.38^{+0.04}_{-0.04}$\\
\hline
Bootes II  & $2.52^{+3.19}_{-1.84}\times 10^{5}$ & $32.99^{+5.01}_{-5.03}$ & $2.51^{+6.05}_{-1.85}$  & $0.16^{+0.02}_{-0.02}$ \\
\hline
Willman I  & $2.90^{+1.59}_{-1.13}\times 10^{5}$ & $19.97^{+4.51}_{-4.12}$ & $13.04^{+17.78}_{-7.28}$  & $0.09^{+0.02}_{-0.02}$ \\
\hline
Coma Berenices  & $1.07^{+4.22}_{-3.46}\times 10^{6}$ & $54.94^{+4.12}_{-4.19}$ & $2.278^{+1.42}_{-0.93}$  & $0.27^{+0.02}_{-0.02}$\\
\hline
\end{tabular}
\end{table}

\section{Bounds on the DM scattering cross section with dSphs stars from {\it Fermi}-LAT data}
\label{sec:fermi_bound}
\begin{figure}[hbt]
    \centering
    \hspace{-2.1cm}
    \includegraphics[scale=0.45]{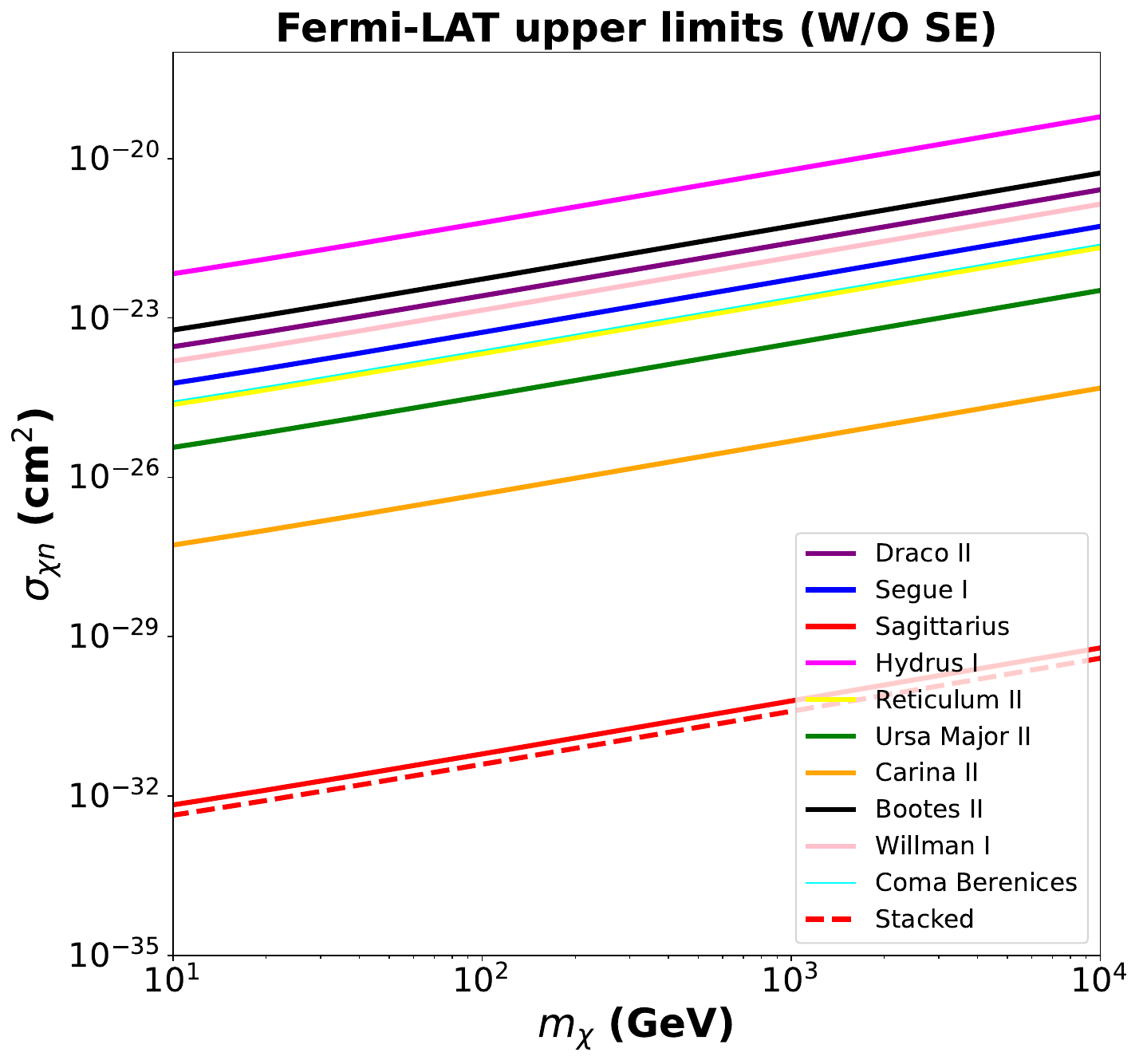}
        \caption{Upper limits on DM mass versus DM-nucleon scattering cross section from individual dSphs and from combined all dSphs (stacked) using the {\it Fermi}-LAT data.}
    \label{fig:bounds_actual}
\end{figure}

\noindent The objective of this paper is to derive the constraints on DM parameter space using {\it Fermi}-LAT flux upper limits from the direction of our selected dSphs. In Sec. \ref{sec:fermi_data} we present the bin-by-bin differential flux upper limits (\fig \ref{fig:diff_flux_fermi}). We perform the binned likelihood analysis by defining the total likelihood function $\mathcal{L}_{i,j}$ for $i{\rm th}$ dSph at the $j{\rm th}$ energy bin which can be written as
\begin{align}
    \label{eq:likehood}
    \mathcal{L}_{i,j} = e^{-N_{\rm exp}}\prod_{j}\frac{{\lambda^{n_j}}_{i,j}}{{n_j}!}\,
\end{align}

\noindent where $n_j$ denotes the measured number of counts while $N_{\rm exp}$ is the total number of expected counts from the source model, including signal and background predictions, and for each dSph the expected count in $j{\rm th}$ bin is labeled as $\lambda_{i,j}$. Now for a given DM mass, we calculate the expected flux from dSphs using \eqs \eqref{eq:dphidE_exp} and \eqref{eq:diff_flux} for box-shaped spectrum and then compare this flux with the {\it Fermi}-LAT upper limits shown in \fig \ref{fig:diff_flux_fermi} following \eq \eqref{eq:dphidE_fermi}. We obtain the bounds on DM annihilation rate $\Gamma_{\rm ann}$ which eventually get translated into the DM parameter space ($m_{\chi}-\sigma_{\chi n}$) in \fig \ref{fig:bounds_actual} corresponding to $R_{\ast}$ and $M_{\ast, \rm tot}$ values of each dSphs (Table~ \ref{tab:dsphs_list}).

\noindent We also perform the stacked analysis using the joint likelihood method ($\mathcal{L}_{\rm joint} = \prod_{i}\mathcal{L}_i$). In \fig \ref{fig:bounds_actual}, the stacked constraints are shown in a dashed red color curve. As expected from \fig \ref{fig:flux_plot}, in \fig \ref{fig:bounds_actual}, we similarly observe that the strongest bound on the DM-nucleon scattering cross section comes from Sgr dSph. 

\noindent We find that the scattering cross section of DM with nucleon can be probed as stringent as $\sim 10^{-33} ~{\rm cm}^2$ from the stacked limits. In our study, we assume the equilibrium hypothesis between the DM capture rate inside the dSphs stellar population and the annihilation rate to obtain the most stringent limits in a conservative approach. Our relatively weak constraints on the scattering cross section compared to other astrophysical sources \cite{Leane:2024bvh} stem from the shallower gravitational potential wells of dSphs. This results in lower DM densities within the stellar bodies, reducing the capture rate and the subsequent annihilation signal. Nonetheless, these galaxies remain invaluable targets due to their low background contamination and the unique insights they provide into DM interactions.

\noindent To improve the bounds on the DM-nucleon scattering cross section, future telescopes with higher sensitivity to gamma rays and broader energy coverage, such as the Cherenkov Telescope Array (CTA) \cite{Hofmann:2023fsn}, will play a pivotal role. These instruments will enable deeper and more precise observations of dSphs, providing critical data to refine our constraints. Additionally, the application of the Sommerfeld enhancement (SE) \cite{Sommerfeld:1931qaf}, which accounts for the velocity-dependent amplification of the DM annihilation rate at low velocities, can offer a promising avenue for tightening our current bounds. This effect is particularly relevant for dSphs, where DM particles are expected to have low velocity dispersions, making them an ideal environment for probing this phenomenon. Following this motivation, in the later section, we examine how the SE can impact and improve our current bounds obtained from this model independent approach.

\section{Effect of astrophysical uncertainties}
\label{sec:uncertainty}
\noindent In this section, we assess the impact of observational uncertainties in key physical parameters on the predicted gamma ray flux and the resulting constraints on the DM–nucleon scattering cross section. This is particularly relevant, as such uncertainties propagate into the inferred dark matter density profiles. We consider the Sgr dSph as a representative case, given that it yields the most stringent bounds on $\sigma_{\chi n}$ among the ten dSphs studied. Although similar analyses can be performed for the remaining targets, Sgr dominates the stacked constraint, with the next most constraining system, Carina II, producing limits that are approximately $\mathcal{O}(4\text{--}5)$ times weaker (see Fig.~\ref{fig:bounds_actual}). We therefore restrict our analysis to Sgr to estimate the potential impact of parameter uncertainties on the final results. This issue is revisited in the discussion section.

\begin{figure}
    \centering

    \begin{subfigure}[b]{0.48\textwidth}
        \centering
        \includegraphics[width=\textwidth, height=7cm]{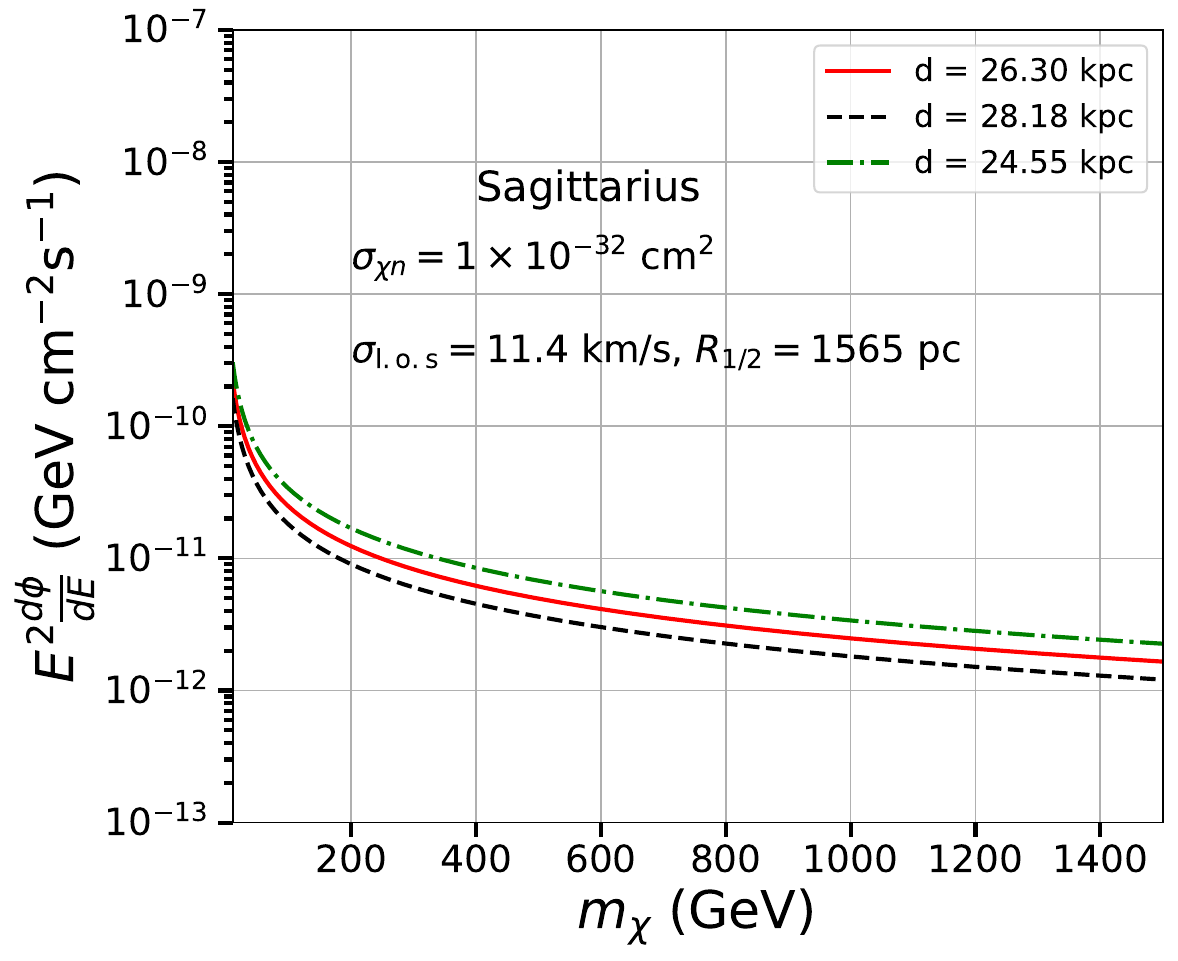}
        \caption{}
    \end{subfigure}
    \hfill
    \begin{subfigure}[b]{0.48\textwidth}
        \centering
        \includegraphics[width=\textwidth, height=7cm]{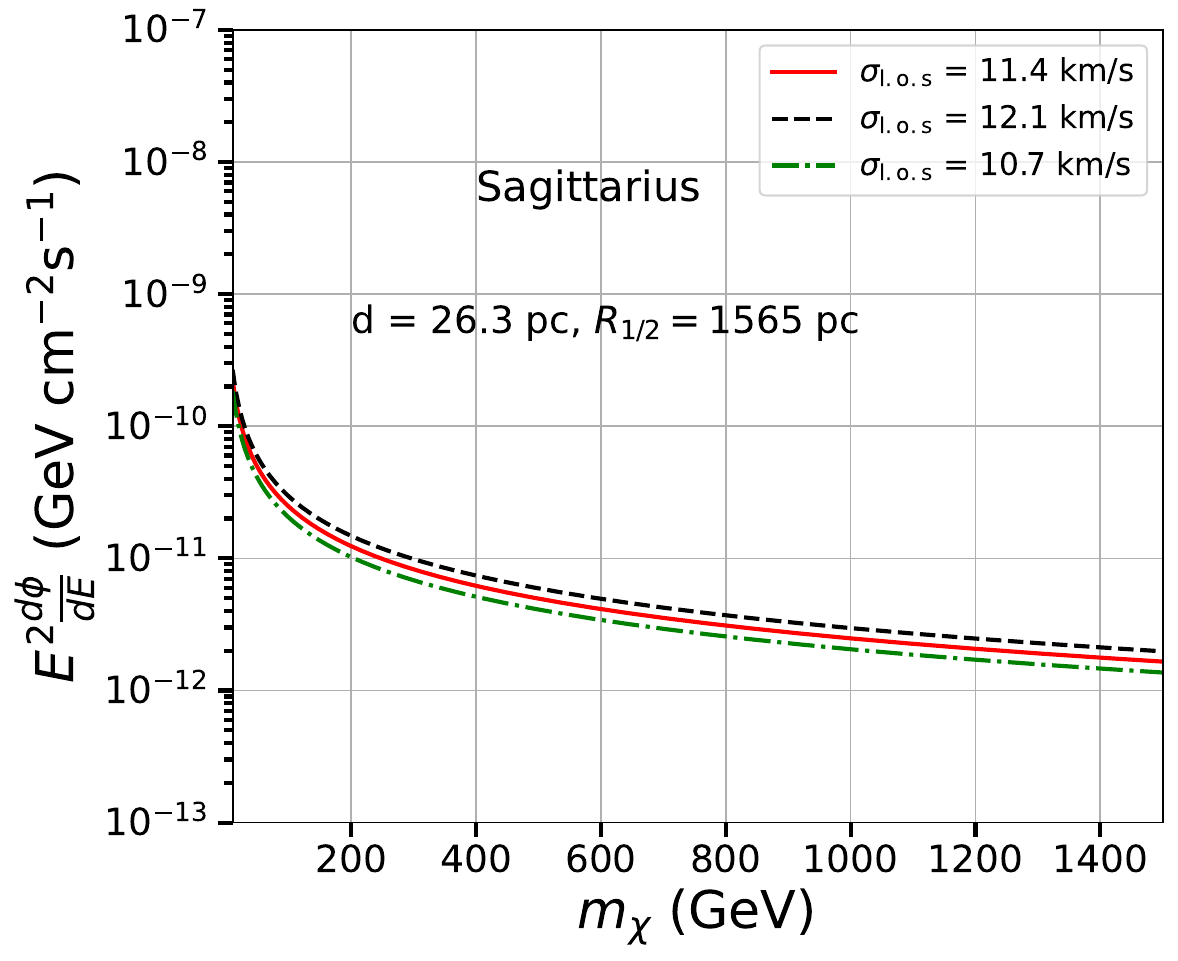}
        \caption{}
    \end{subfigure}

    \vspace{0.5cm} 

    \begin{subfigure}[b]{0.48\textwidth}
        \centering
        \includegraphics[width=\textwidth, height=7cm]{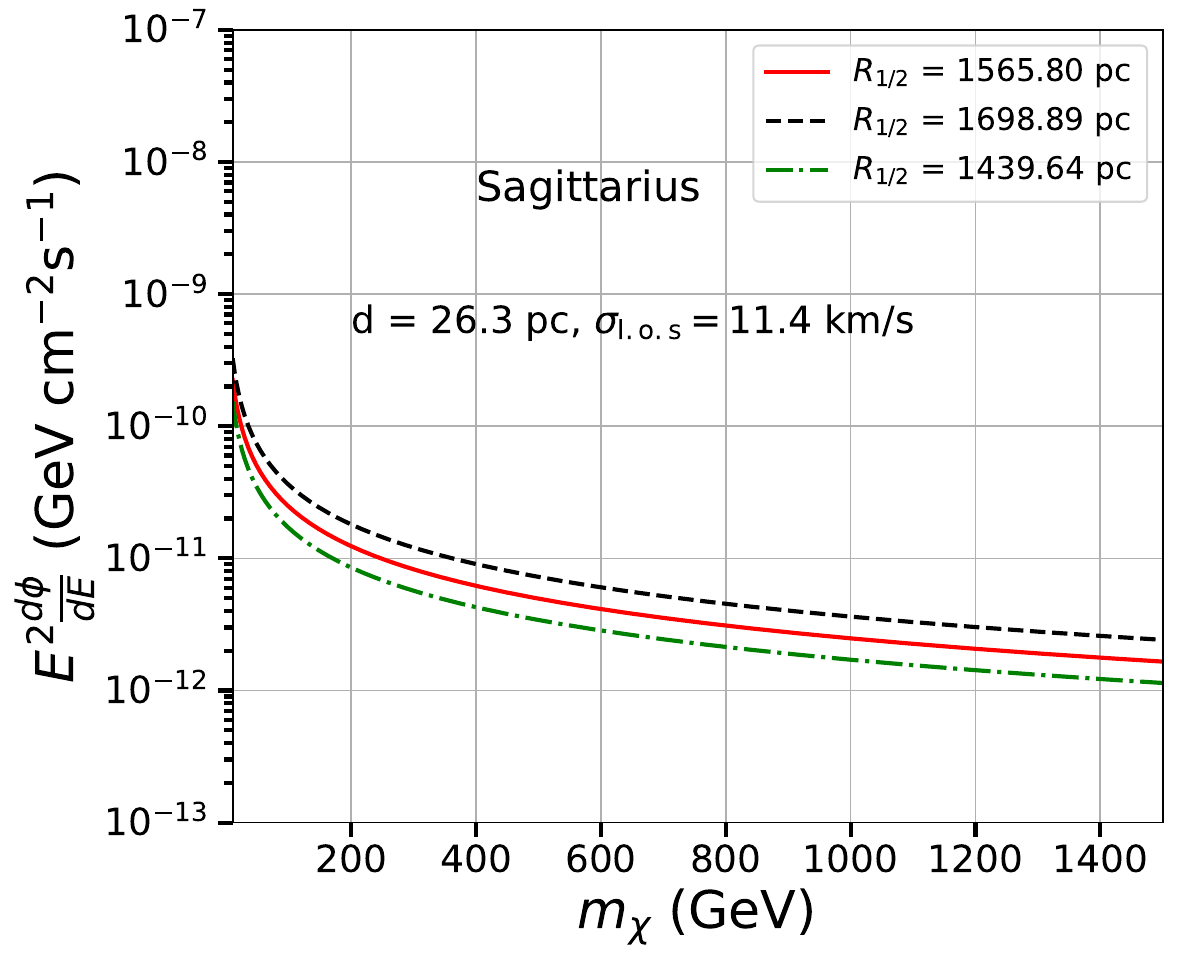}
        \caption{}
    \end{subfigure}
    \hfill
    \begin{subfigure}[b]{0.48\textwidth}
        \centering
        \includegraphics[width=\textwidth, height=7cm]{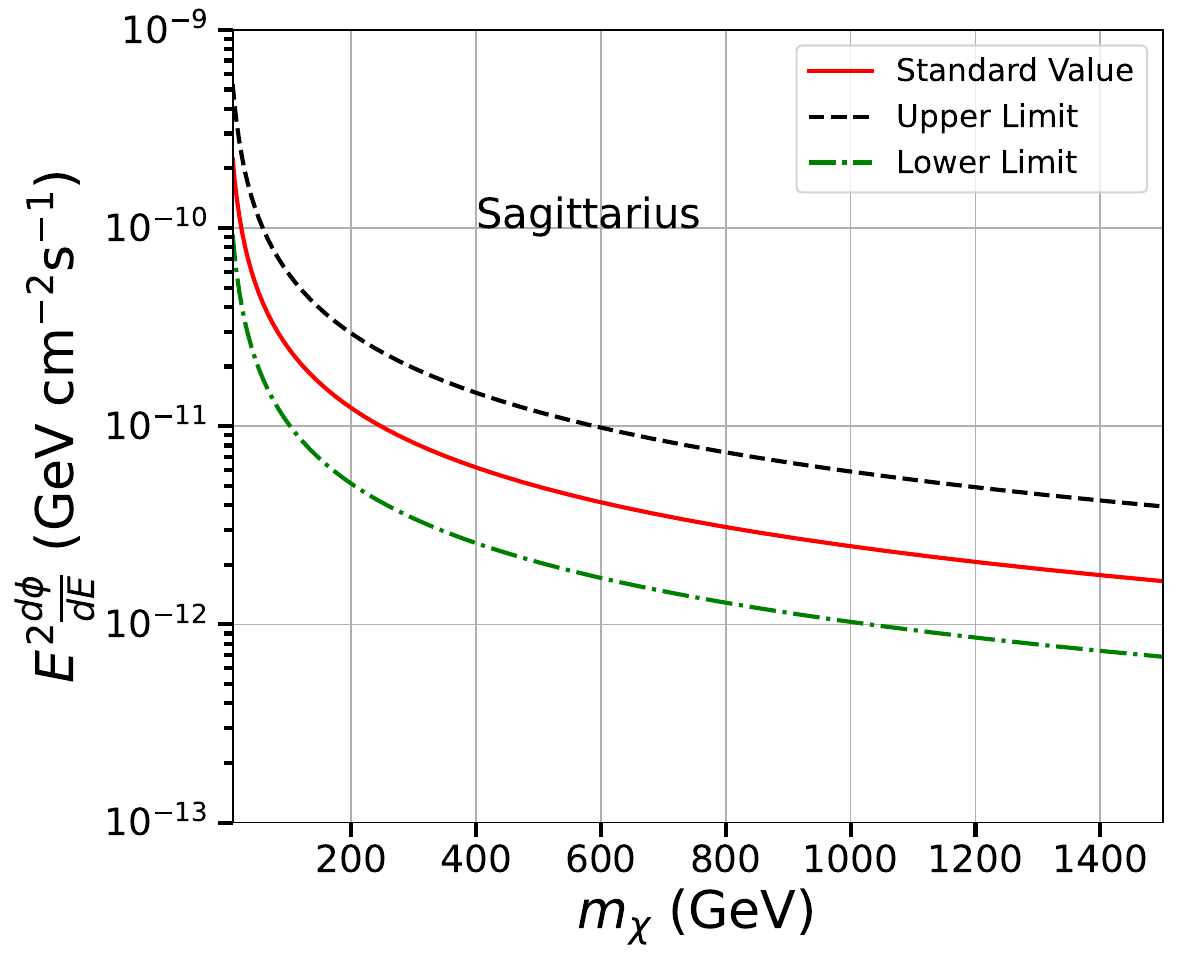}
        \caption{}
    \end{subfigure}

    \caption{Effect of individual and combined astrophysical uncertainties on DM annihilation fluxes from Sgr dSph. The uncertainty on the fluxes can originate from the observational uncertainties on the measurement of dSph parameters (a) d, (b) $\sigma_{\rm l.o.s}$, and (c) $R_{1/2}$. The variations on these dSph parameter values will affect the $r_s$ and $\rho_s$ values used in the dSph DM profile (\eq \ref{eq:nfw}). }   
    \label{fig:uncertain}
\end{figure}

\begin{figure}[hbt]
 \hspace{-1.2cm}
    \includegraphics[scale=0.65]{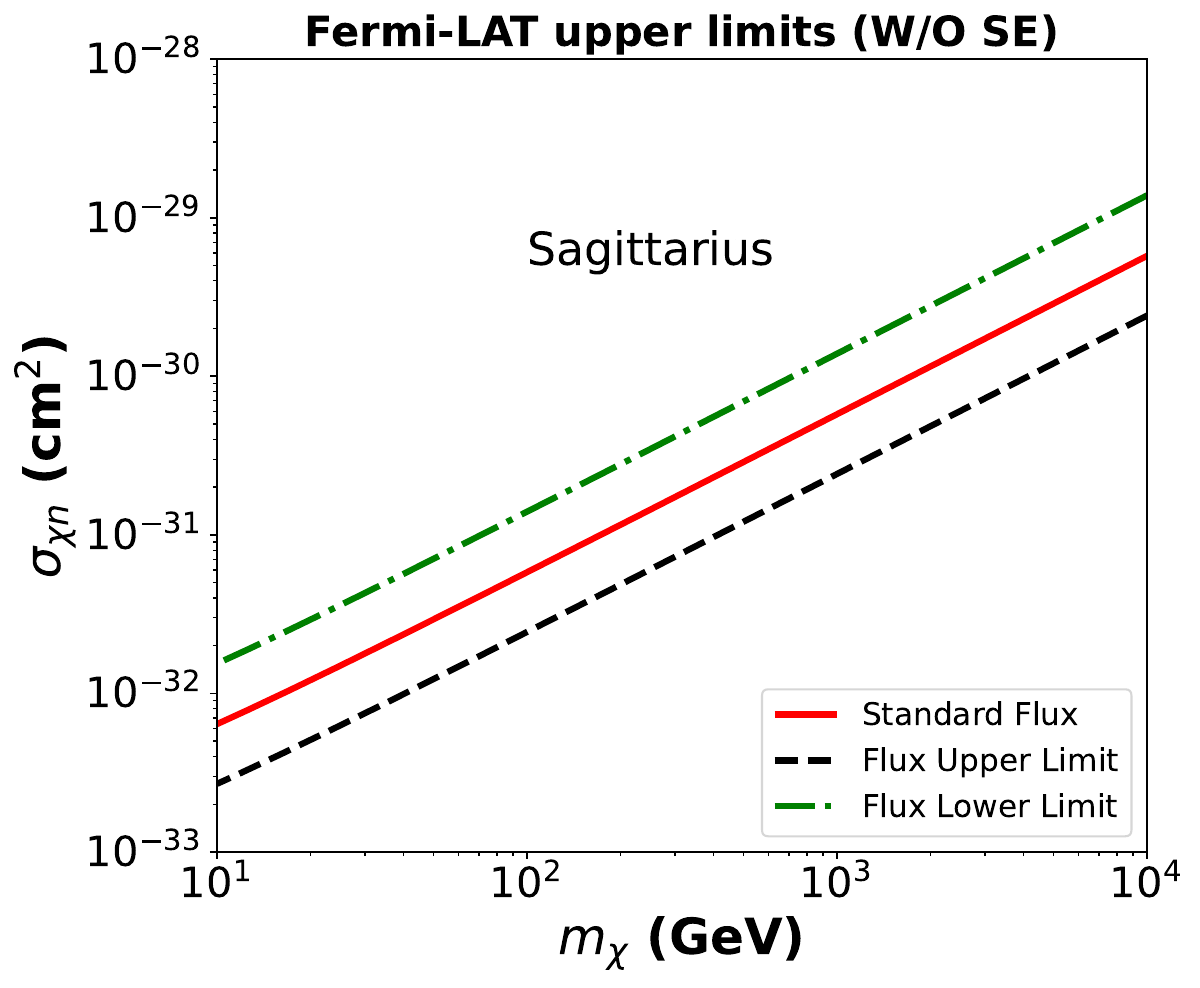}
    \caption{Effect of astrophysical uncertainties on the DM-nucleon scattering cross section bounds from Sgr dSph. The black (green) limits correspond to the case where upper (lower) flux values are used, considering the uncertainties ($1\sigma$) in dSph parameters.} 
    \label{fig:bound_uncertain}
\end{figure}

\noindent From \eqs \ref{eq:nfw} and \ref{eq:dphidE_exp}, we understand that the gamma-ray fluxes from DM annihilation are subject to uncertainties on account of the errors in determining the astrophysical parameters $d$ (heliocentric distance), $\rho_s$ (characteristic density) and $r_s$ (scale radius). The values of $\rho_s$ and $r_s$ are computed from analytical expressions given in Refs.~\cite{Bhattacharjee:2020phk, Evans:2016xwx} which in turn depend upon $R_{1/2}$ (half-light radius) and $\sigma_{\rm l.o.s}$ (velocity-dispersion). Thus, the main source of uncertainties in the flux comes from ambiguity in the determination of $d, R_{1/2}$, and $\sigma_{\rm l.o.s}$. Using the central values for these dSph parameters, the bounds on $\sigma_{\chi n}$ are displayed in \fig \ref{fig:bounds_actual}. Here, we first study the effect of individual uncertainty by varying one parameter at a time. To this end, in \fig \ref{fig:uncertain}, we present the possible uncertainties in the DM-induced gamma ray fluxes as a consequence of ambiguities in $d$ (\fig \ref{fig:uncertain} (a)), $\sigma_{\rm l.o.s}$ (\fig \ref{fig:uncertain} (b)), $R_{1/2}$ (\fig \ref{fig:uncertain} (c)), and combining all three contributions (\fig \ref{fig:uncertain} (d)). We fix $\sigma_{\chi n} = 1\times 10^{-32}$ cm$^2$ for all the plots in \fig \ref{fig:uncertain} and the relevant parameters of Sgr dSph are varied within their $1\sigma$ uncertainties mentioned in Tables \ref{tab:dsphs_list} and \ref{tab:list_nfw}. In all plots of \fig \ref{fig:uncertain}, the standard gamma ray fluxes (shown in red colour curves) are computed using the central values of the dSph parameters. For simplicity and in the absence of adequate information, we assume no correlations among these three parameters. We notice from \fig \ref{fig:uncertain} (b) that the error in the dispersion velocity $\sigma_{\rm l.o.s}$ has the least effect on the gamma-ray fluxes, while the actual uncertainty band, which is a combination of all three contributions and presented in \fig \ref{fig:uncertain} (d), shows $5-8$ times uncertainty in the fluxes from the central values (red colour curve). It should also be noted that, due to the large relative error in $R_{1/2}$ measurement, this parameter contributes more to the flux uncertainty and hence should be measured more precisely. Finally, we propagate the uncertainty in the DM-induced gamma-ray fluxes into the DM-nucleon scattering cross section $\sigma_{\chi n}$ and the result is displayed in \fig \ref{fig:bound_uncertain} in $\sigma_{\chi n} - m_\chi$ plane for Sgr dSph. The bounds shown in red colour correspond to the case where central values of dSph parameters are used, while the band between green and black colour represents the maximum possible uncertainty in the constraints for Sgr dSph. From \fig \ref{fig:bound_uncertain} we observe that the maximum error in the bounds on $\sigma_{\chi n}$ can be of $\mathcal{O}(1)$.


\section{Effect of Sommerfeld Enhancement on scattering cross section upper limits}
\label{sec:sommerfeld}
\noindent In this section, we study a scenario where DM annihilation occurs through LLLMs, incorporating the Sommerfeld enhancement (SE), first introduced by Arnold Sommerfeld in 1931~\cite{Sommerfeld:1931qaf}, into the DM annihilation cross section. This effect refers to the fact that in the presence of long-range attractive potential, the annihilation cross section of nonrelativistic DM particles is significantly enhanced, leading to a larger observational signal. The Sommerfeld effect is more pronounced at low relative velocities (Sommerfeld factor $ \propto 1/{\rm velocity}$) and dSphs are considered some of the best targets to observe such phenomena because the relative velocities of DM particles in dSphs tend to be much lower than other astrophysical objects such as galaxy clusters or Milky Way~\cite{Boddy:2017vpe, Ando:2021jvn}.\\
\noindent
It should be noted that the effects of the SE depend upon the nature of the force carrier~\cite{Liu:2013vha, Arkani-Hamed:2008hhe} as the annihilation cross section, and hence, the Sommerfeld factor would be different for different force carriers. In Refs.~\cite{Dasgupta:2012bd, Du:2021cmt, Feng:2015hja, Bell:2021pyy}, for instance, the mediator is a vector gauge boson (dark photon model) which gives neutrino as DM signal. However, the decay of a vector boson directly into two gamma ray photons is forbidden~\cite{Bell:2021pyy}, i.e., if $\phi$ is a dark photon, $\phi \rightarrow\gamma\gamma $ is prohibited. We, therefore, adopt a very simple model in order to get gamma ray flux from the decay of the mediator $\phi$ along with the inclusion of the SE. \\
\noindent For the DM model considered here, we assume that the DM particles experience a long-range force mediated by an LLLM, $\phi$, before they undergo self-annihilation into the same mediator $\phi$ which later decays to produce a detectable gamma ray signal. So, in this very simplified model, we have the same mediator $\phi$ that induces the Sommerfeld effect as well as provides a box-shaped spectrum due to DM annihilation via LLLM. Under this assumption, the SE-induced annihilation cross section rate ($\langle\sigma_{\rm ann} v\rangle_S$) can be expressed as the product of the Sommerfeld factor and the Born-approximated annihilation cross section ($\langle\sigma_{\rm ann} v\rangle_{\rm Born}$), as shown below~\cite{Lu:2017jrh},
\begin{align}
    \label{eq:sigma_v_som}
    \langle\sigma_{\rm ann} v\rangle_S = \langle\sigma_{\rm ann} v\rangle_{\rm Born}\langle S_{\rm swave}\rangle \, ,
\end{align}
where $\langle S_{\rm swave}\rangle$ denotes the thermally averaged SE factor for s-wave and we consider $\phi$ as a scalar particle that may eventually decay into two gamma photons. Another important assumption we make to maximize the gamma ray flux is that the DM particle is also a scalar \cite{Chowdhury:2016mtl, Bell:2024uah} \footnote{Although we have focused on the scalar DM case for SE, a fermionic scenario can also be considered; see, e.g.,~\cite{Phoroutan-Mehr:2024cwd, Hisano:2004ds, Biondini:2018pwp}. However, the main issue is that, for a fermion, $\chi\chi\to \phi\phi$ annihilation process suffers from p-wave suppression reducing the gamma ray signal from dSphs. We plan to consider such cases in future work.}.  In that case, the thermally averaged annihilation cross section at the tree level (for scalar DM interacting with a scalar mediator) is given by~\cite{Liu:2014cma}
 
\begin{align}
    \label{eq:sigma_v_scalar_new}
\langle\sigma_{\rm ann} v\rangle_{\rm Born} = \frac{\pi \alpha_\chi^2\varepsilon^4_\phi}{4 m_\chi^2}\frac{\sqrt{1-\varepsilon^2_\phi}}{(1-\varepsilon^2_\phi/2)^2}\,,    
\end{align}

\noindent where $\varepsilon_\phi = m_\phi/m_\chi$ and $\alpha_\chi = g_\chi^2/4\pi$ represent the dark fine structure constant with $g_\chi$ being the gauge coupling of the long-range force mediated by $\phi$. The enhancement factor for s-wave thermally averaged annihilation can be expressed as~\cite{Cassel:2009wt, Feng:2010zp}    
\begin{align}
    \label{eq:som_average}
    \langle S_{\rm swave}\rangle =  \int \frac{S_{\rm swave} (e^{-v^2/2v_0^2})}{(2\pi v_0^2)^{3/2}} d^3v\,,
\end{align}

\noindent where the DM velocity inside the object is $v_0 = \sqrt{2~T_{\star,c}/m_\chi}$. In our dSphs analysis, we use $T_{\star,c}\sim 30000 K \sim 2.59\times 10^{-9}$ GeV~\cite{Mashchenko:2005cs}. The s-wave SE factor ($S_{\rm swave}$) can be derived analytically by approximating the Yukawa-type long-range interaction for nonzero mediator mass to the Hulth\'{e}n potential and is given by~\cite{Slatyer:2009vg, Cassel:2009wt, Feng:2010zp, Liu:2013vha}

\begin{align}
    \label{eq:som}
    S_{\rm swave} = \frac{\pi}{\beta}\frac{\sinh{2\pi \beta \zeta}}{\cosh{2\pi \beta \zeta} - \cos({2\pi\sqrt{\zeta-\beta^2\zeta^2}})}\,,
\end{align}
\noindent where $\beta = v/(2\alpha_\chi)$ and $\zeta = 6\alpha_\chi m_\chi/(\pi^2 m_\phi)$. \\

\begin{figure}[hbt]
 \hspace{-1.2cm}
    \includegraphics[scale=0.35]{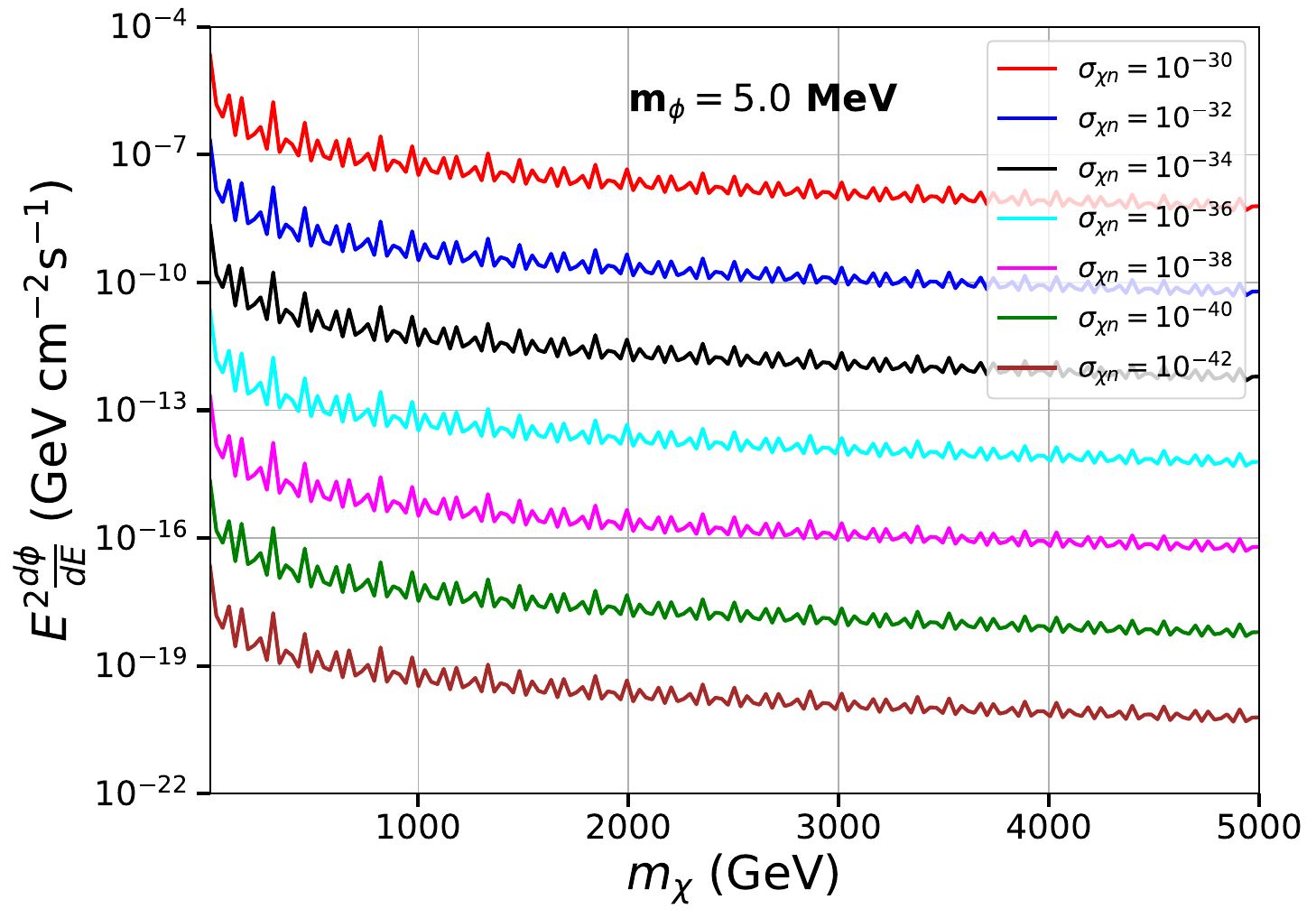}
    \includegraphics[scale=0.35]{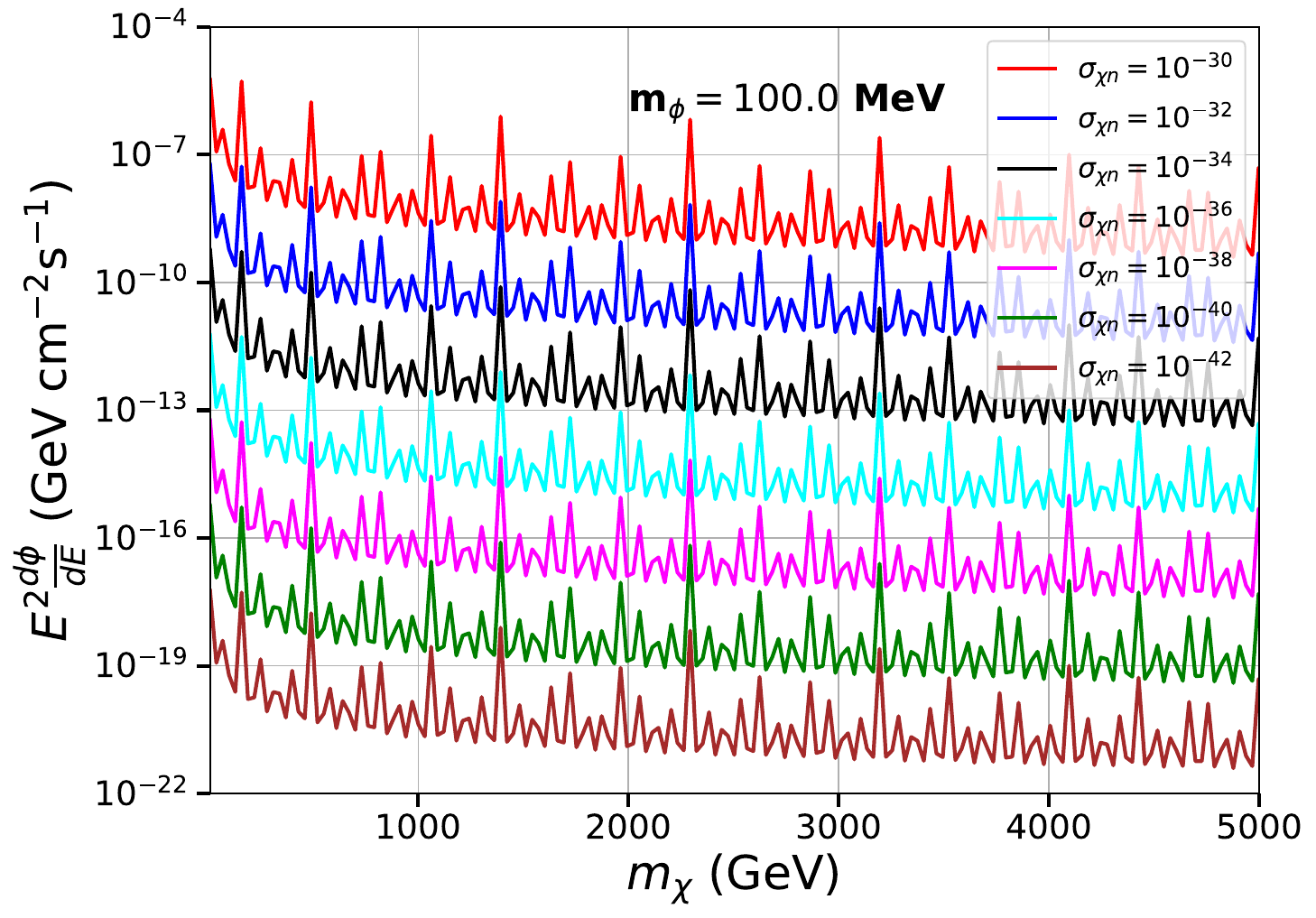}
    \caption{Effect of SE on DM annihilation flux from Sgr dSph. The left (right) plot corresponds to the mediator mass $m_\phi = 5.0~(100.0)$ MeV. The rapid oscillation in the flux is a distinctive feature of the Sommerfeld effect owing to the formation of bound states near the threshold in the presence of attractive potential~\cite{Acevedo:2024ava, Ando:2021jvn, Lattanzi:2008qa, Hisano:2003ec}.}   
    \label{fig:sommer_flux}
\end{figure}

\noindent We compute the total annihilation rate, i.e. $\Gamma_{\rm ann}$ and thereby the annihilation flux of gamma rays in the presence of SE by employing \eqs from \ref{eq:sigma_v_som} to \ref{eq:som} and \eq \ref{eq:Ann_rate}. Next, we determine the value of $\alpha_\chi$  by fixing the annihilation rate at freeze-out as $\langle\sigma_{\rm ann} v\rangle = 2.2~\times~10^{-26}~{\rm cm}^3/{\rm s}$~\cite{1987ApJ...321..560G} to satisfy the DM thermal relic density $\Omega_\chi h^2 = 0.12$ which implies
\begin{align}
    \alpha_\chi = \left(\dfrac{0.097}{\varepsilon_\phi^2}\right)\left(\dfrac{m_\chi}{\rm TeV}\right)\,.
\end{align}

\begin{figure}[hbt]
    \hspace{-1.2cm}
    \includegraphics[scale=0.35]{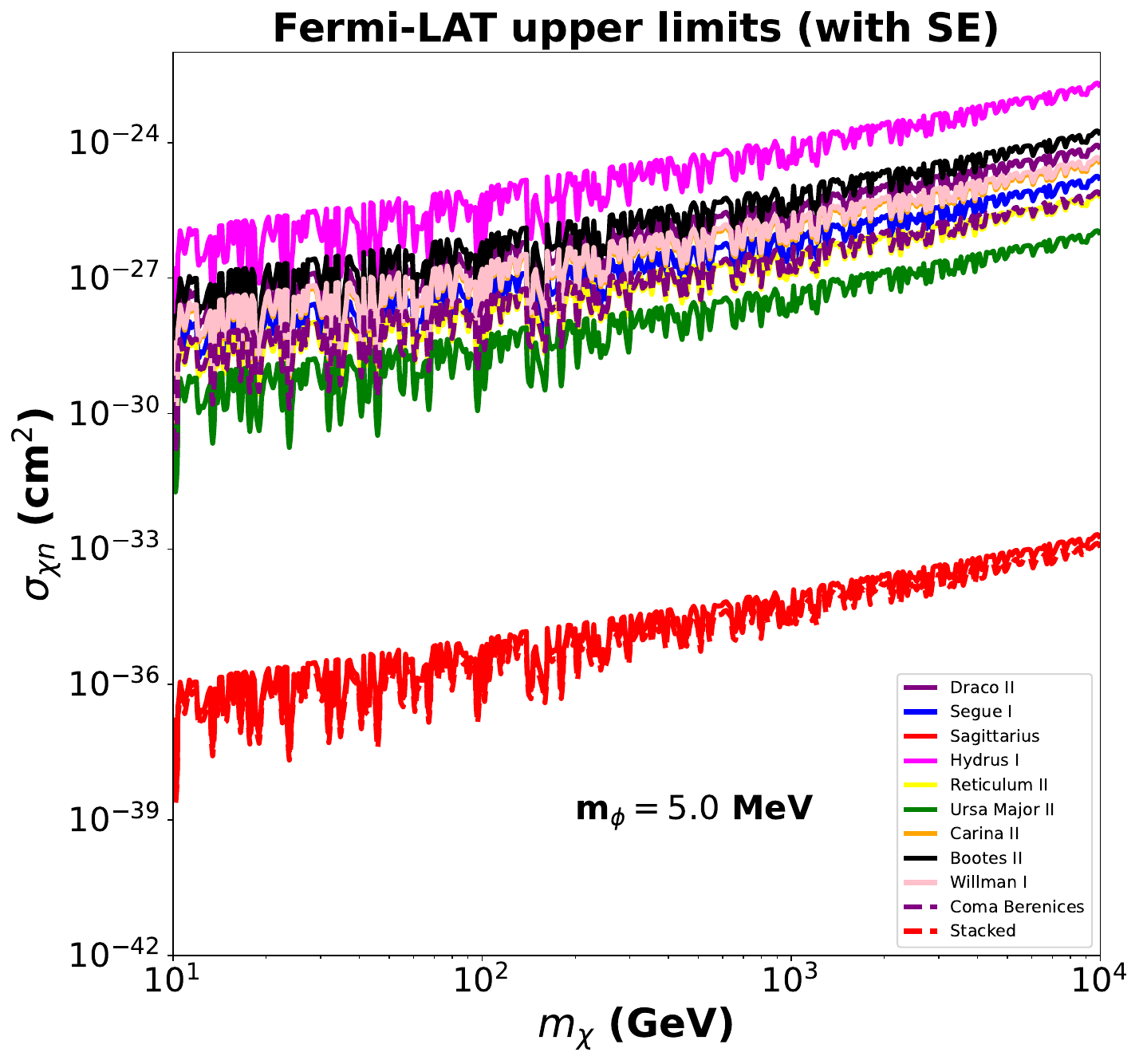}
    \includegraphics[scale=0.35]{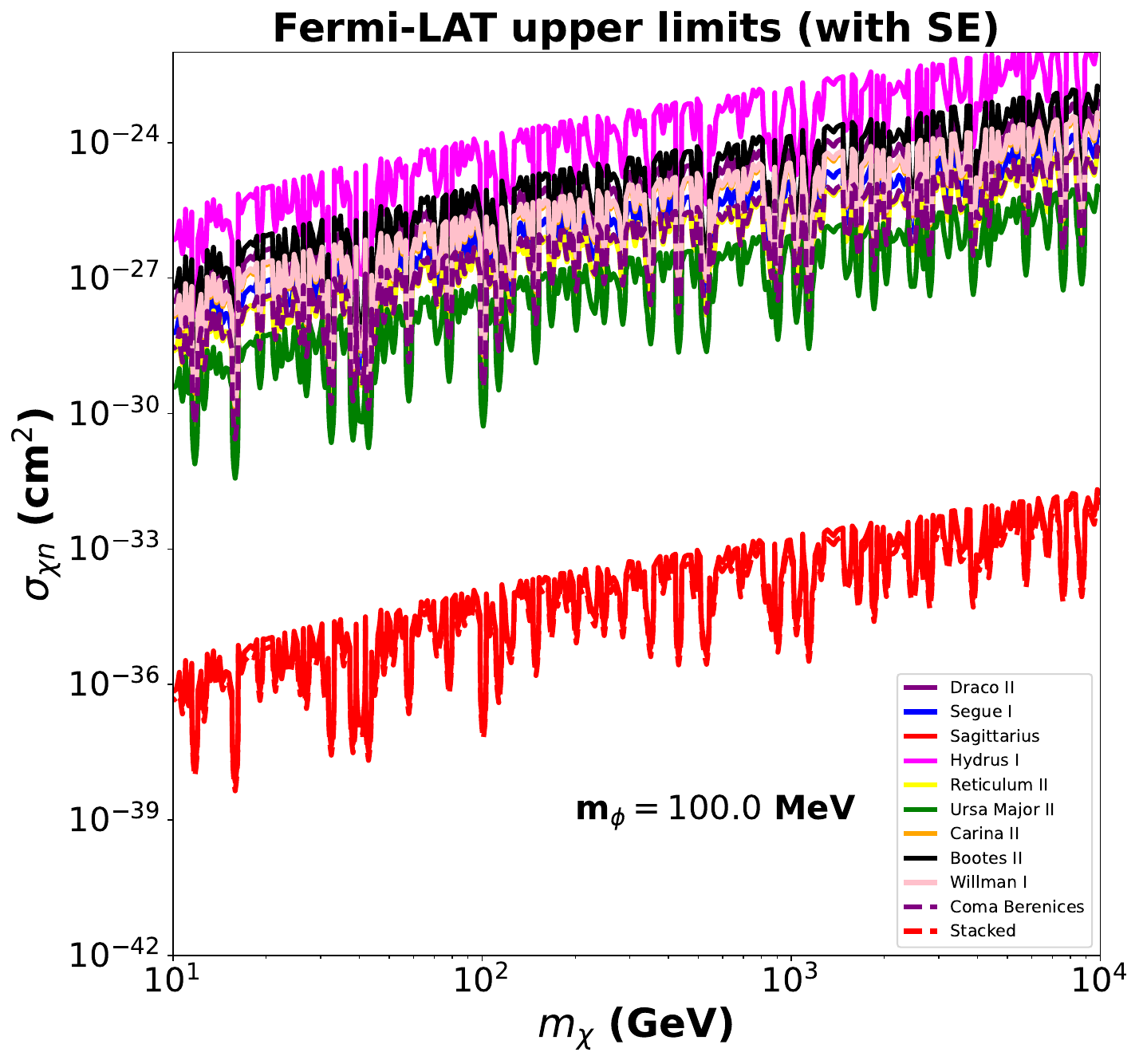}
        \caption{Upper limits on DM-nucleon scattering cross section as a function of DM mass including the SE using the {\it Fermi}-LAT data. The left (right) plot corresponds to the mediator mass $m_\phi = 5.0 ~(100.0)$ MeV.}
    \label{fig:bounds_actual_SE}
\end{figure}

\noindent In \fig \ref{fig:sommer_flux} we demonstrate the Sommerfeld enhanced gamma ray flux for Sgr dSph as a function of DM mass for different values of scattering cross section, $\sigma_{\chi n}$. The left (right) plot corresponds to the mediator mass $m_\phi = 5.0~ (100.0)$ MeV. The wiggles in \fig \ref{fig:sommer_flux} represent the usual nature of the SE. We observe that the flux is getting significantly enhanced due to the Sommerfeld effect (compare \fig \ref{fig:flux_plot} and \fig \ref{fig:sommer_flux}). Moreover, we also notice a slight dependence of gamma ray flux on the mediator mass $m_\phi$. We, then compute the bounds on DM scattering cross section as a function of DM mass following the same approach mentioned in Sec. \ref{sec:fermi_bound} using {\it Fermi}-LAT upper limits derived in \fig \ref{fig:diff_flux_fermi}. The results are furnished in \fig \ref{fig:bounds_actual_SE} for both individual dSph and stacked analyses. \\

\noindent From Fig.~\ref{fig:bounds_actual_SE}, it is evident that the inclusion of the SE significantly strengthens the constraints, improving them by nearly 3 to 4 orders of magnitude compared to the case without SE (Fig.~\ref{fig:bounds_actual}). The current limits reach $\sim~10^{-36}~{\rm cm}^2$ for DM masses around 100 GeV. This happens due to the presence of long-range attractive interaction of DM particles mediated by  LLLMs, the equilibrium timescale reduces significantly, increasing the DM annihilation flux to its maximum value. It is to be noted that with the increase of the mediator mass from $m_\phi$ the bounds become better to some extent. Similar observations have been shown in \fig \ref{fig:sommer_flux} of Ref.~\cite{Feng:2015hja}. We also notice the rapid oscillations in \fig \ref{fig:bounds_actual_SE} which is a well known feature of the SE due to the formation of bound (or resonancelike) states. \\

\begin{figure}[hbt]
    \centering
    \hspace{-1.2cm}
    \includegraphics[scale=0.50]{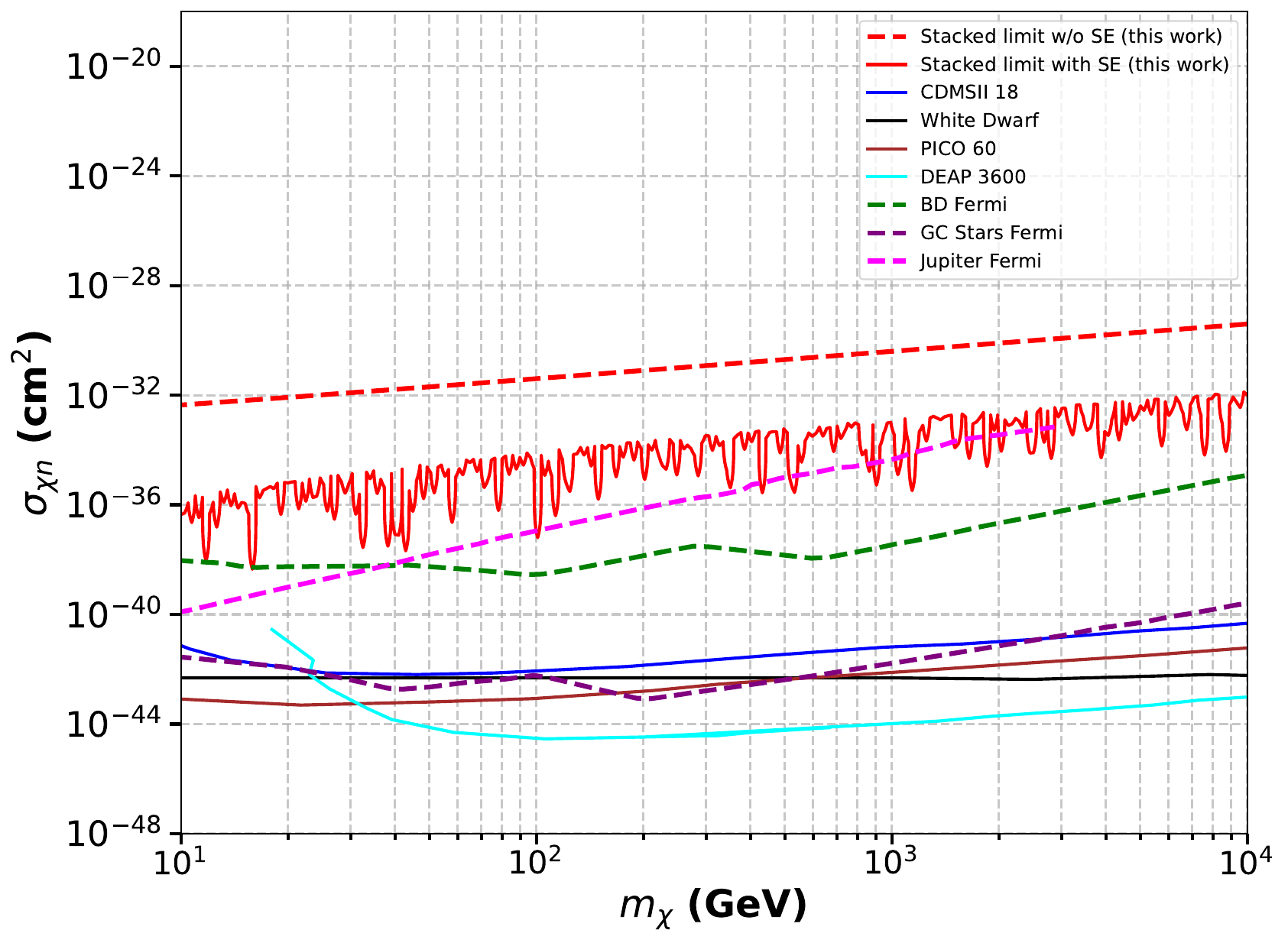}
        \caption{Comparison between the bounds obtained from dSphs in this work with those already available in the literature. We also display the direct detection constraints on spin-independent DM-nucleon cross sections from different experiments such as CDMS II~\cite{SuperCDMS:2017mbc}, PICO 60 \cite{PICO:2019vsc}, and DEAP 3600 \cite{DEAP:2019yzn}.  and the indirect detection bounds from other astrophysical objects are adapted from Refs.~\cite{Leane:2021ihh, Leane:2021tjj, Acevedo:2023xnu, Leane:2024bvh}.}
    \label{fig:compare_bounds}
\end{figure}

\noindent In \fig \ref{fig:compare_bounds}, we compare our results, both with SE and without SE, to the limits reported by various direct detection \cite{SuperCDMS:2017mbc, PICO:2019vsc, DEAP:2019yzn} and astrophysical observations \cite{Leane:2021ihh, Leane:2021tjj, Acevedo:2023xnu, Leane:2024bvh}.  Our limits are far above those reported by underground direct detection experiments by CDMS II~\cite{SuperCDMS:2017mbc}, PICO 60 \cite{PICO:2019vsc}, and DEAP 3600 \cite{DEAP:2019yzn}. Moreover, we also plot the limits from other astrophysical studies performed on Galactic center (GC) stars \cite{Leane:2024bvh}, GC population of brown dwarfs (BDs) \cite{Leane:2021ihh}, white dwarfs (WDs) \cite{Acevedo:2023xnu}, and Jupiter \cite{Leane:2021tjj}. We notice that for DM mass around 10-20 GeV, our limits are comparable to those obtained from BDs, and for higher DM masses ( $ > \sim 500$ GeV) stacked dSphs bounds are stronger than Jupiter. Although the resulting bounds on DM-nucleon interactions derived from dSphs are weaker compared to those obtained from direct detection experiments or more massive compact objects like BDs and WDs, they offer the advantage of being cleaner, and relatively free from significant astrophysical backgrounds. Furthermore, investigating the DM capture rate within the stellar populations of dSphs complements traditional indirect DM searches in these galaxies \cite{VERITAS:2024usn, McDaniel:2023bju}, thereby enhancing the comprehensive exploration of DM properties.

\vspace{1cm}

\section{Conclusion and Discussion}
\label{sec:conclusion}

\noindent This paper investigates the DM signal from nearby dSphs ($<$ 50 kpc distance), considering a phenomenological framework in which DM particles first capture within the stars inside dSph and subsequently annihilate into LLL mediators. The mediator will later decay into gamma rays that can be detected by gamma ray telescopes such as {\it Fermi}-LAT. For this, we have analyzed nearly 16 years of data from {\it Fermi}-LAT to look for the gamma-ray emission from the selected dSphs and present the observed limits in \fig \ref{fig:diff_flux_fermi}. In the absence of any excess signal, we estimate the $95\%$ CL conservative upper limits on DM-nucleon scattering cross section ($\sigma_{\chi n}$) as a function of DM mass ($m_\chi)$.\\

\noindent We find that Sgr dSph provides the best limits among all other dSphs considered in this work, and the stacked limits can be as low as $\sim 10^{-33}$ cm$^2$. Since our goal is to estimate the maximum possible upper limits expected from dSphs while exploring the two-step (cascade) process of DM annihilation into gamma photons, we assume that equilibrium is achieved between DM capture and annihilation. This maximizes the rate at which captured DM annihilates into mediators, making our bounds conservative. However, the equilibrium timescale is somewhat larger than the actual age of the dSphs, and the bounds might get weakened if the above-mentioned assumption is relaxed. \\

\noindent In the later part of this work, a particular model has been considered where both DM and the mediator are scalar particles. This further allows us to explore an interesting phenomenon, SE, in which, owing to the presence of long-range attractive potential sourced by the scalar mediators, the DM annihilation cross section and, hence, the detection prospects of such DM candidates enhance significantly. This is particularly important when the DM dispersion velocity is low, as in the case of dSphs. We show the resonancelike features of the SE in the DM-induced gamma ray flux in \fig \ref{fig:sommer_flux} that are also reported in~\cite{Feng:2015hja}. We notice that the effect of the mediator mass, $m_\phi$, on the gamma ray flux is not very significant, and it changes only by a factor of a few. We then derive the upper limits on DM-nucleon scattering cross section in the presence of SE following a similar approach as mentioned in Sec. \ref{sec:fermi_bound}. As expected, the limits are significantly improved for Sgr dSph (and also for the stacked limit) with the inclusion of the SE. \\

\noindent Our study also highlights the impact of observational uncertainties in key astrophysical parameters on the gamma-ray flux from DM annihilation in dSphs. Using Sgr as a representative case, we find that variations, especially in the half-light radius $R_{1/2}$, can lead to significant shifts in the derived limits on the DM–nucleon scattering cross section, with uncertainties reaching up to $\mathcal{O}(1)$. This underscores the need for more precise kinematic and structural measurements of dSphs to strengthen indirect DM searches. Although the uncertainty analysis was performed for the case without SE, similar conclusions are expected in its presence, as the SE factor $<S_{\rm swave}>$ (see \eq \ref{eq:som_average}) does not depend on the dSph parameters $d$, $R_{1/2}$, or $\sigma_{\rm l.o.s}$, and is thus largely free from astrophysical uncertainties. Therefore, we expect an average error of $\mathcal{O}(1)$ in the limits on $\sigma_{\chi n}$ for both SE and no SE cases.

\noindent Finally, the bounds on the DM-nucleon cross section derived in this work are compared to those available in other studies as illustrated in \fig \ref{fig:compare_bounds}.
The constraints shown in \fig \ref{fig:compare_bounds} remain much weaker than those reported by underground direct detection experiments. Even in the presence of SE, the sensitivity obtained in the present analysis is several orders of magnitude above the current best experimental limits (say at 100 GeV, $\sigma_{\chi n} < 10^{-46}~{\rm cm}^2$~\cite{LZ:2022lsv, PandaX:2024qfu}). On comparison with the bounds from other celestial objects, we see that the stacked dSphs upper limits (in the presence of SE) are stronger than Jupiter~\cite{Leane:2021tjj} for the DM masses above $\sim 500$ GeV and comparable to the BDs limits~\cite{Leane:2021ihh} for masses in the 10-20 GeV range. We also note that the direct detection bounds from various experiments represent the model independent limits whereas the Sommerfeld enhanced constraints on $\sigma_{\chi n}$ derived in this work are model dependent. Other model independent bounds on the DM parameter space come from the cosmic structure formation which mainly rely on DM-nucleon scattering and not on the details of the particle nature of DM. A few related works on the study of DM interactions from structures formation are; limits from Lyman-alpha forest observations~\cite{Rogers:2021byl}, abundance of Milky Way satellite galaxies~\cite{Nadler:2019zrb,Maamari:2020aqz,DES:2020fxi}, and cosmic microwave background (CMB) measurements~\cite{Gluscevic:2017ywp,Boddy:2018kfv}. However, all of these bounds are more than 2 orders of magnitude weaker in the GeV-TeV mass range than the ones obtained in this work. Indeed the structure formation limits on $\sigma_{\chi n}$ are very significant for the light (sub-GeV) DM scenarios where direct detection experiments have limited sensitivity. In this regard, we want to highlight that, even though dSphs are weaker than other current studies to capture DM, they provide a relatively cleaner environment due to the lack of gas and minimal astrophysical backgrounds. These results contribute to a broader understanding of DM properties, offering constraints that avoid some of the systematic uncertainties inherent in direct detection experiments and complement existing astrophysical bounds. Together, these findings underline the value of multifaceted approaches to probing the elusive nature of DM and its interactions. \\

\noindent To the best of our knowledge, this study represents one of the first attempts to probe the stellar component of dSphs to investigate the DM capture rate. By adopting a dual approach of both model independent and model dependent (incorporating the SE effect), our analysis offers a comprehensive framework for constraining the DM nucleon scattering cross section. This novel perspective sheds light on the potential role of stellar bodies within dSphs as complementary probes of DM interactions, emphasizing their significance alongside other astrophysical sources.\\

\noindent Our study also points to a new and alternative avenue for exploring DM phenomena in dSphs, particularly by highlighting the importance of incorporating velocity-dependent effects, such as the SE, to refine existing bounds. Future telescopes with enhanced sensitivity, such as the CTA, and advancements in stellar modeling are expected to provide critical improvements in the accuracy of DM capture rate estimations. Furthermore, the framework established in this work underscores the value of low-velocity environments in DM studies, offering a fresh perspective on the interplay between stellar populations and DM annihilation. These insights pave the way for a more holistic understanding of DM behavior across diverse astrophysical settings.

\section*{ACKNOWLEDGMENTS}
\noindent We thank Amit Dutta Banik for the important and detailed discussions on Sommerfeld enhancement and constructive suggestions on our work. We also thank Ranjan Laha, Anirban Das, and Satyanarayan Mukhopadhyay for the useful discussions. P.B. acknowledges support from the COFUND action of Horizon Europe’s Marie Sklodowska-Curie Actions research program, Grant Agreement No. 101081355 (SMASH). A.G. further wishes to acknowledge the organizers of the ``Trends in Astro-particle and Particle Physics (TAPP 2024)'' at the Institute of Mathematical Sciences, Chennai, India, during 25th to 27th September 2024 and the ``XXVI DAE-BRNS High Energy Physics (HEP) Symposium 2024'' at Banaras Hindu University (BHU), Varanasi, India, during 19th to 23rd December 2024, for providing an opportunity to present the preliminary results from this work.   


\bibliographystyle{JHEP}
\bibliography{references.bib}

\end{document}